\pdfoutput=1
\RequirePackage{ifpdf}
\ifpdf 
\documentclass[pdftex]{sigma}
\else
\documentclass{sigma}
\fi

\numberwithin{equation}{section}

\newtheorem{Theorem}{Theorem}[section]

\newtheorem{Proposition}[Theorem]{Proposition}
 { \theoremstyle{definition}
\newtheorem{Definition}[Theorem]{Definition}

\newtheorem{Example}[Theorem]{Example}
\newtheorem{Remark}[Theorem]{Remark} }

\usepackage{bm}
\usepackage{dsfont}

\newcommand{\B}{\mathcal{B}}
\newcommand{\K}{\mathcal{K}}
\newcommand{\R}{\mathbb{R}}
\newcommand{\D}{\mathcal{D}}
\newcommand{\bl}{\boldsymbol{\lambda}}
\newcommand{\bmu}{\boldsymbol{\mu}}
\newcommand{\J}{{\rm J}}
\newcommand{\n}{^{(n)}}
\newcommand{\ox}{\overline{x}}
\newcommand{\oy}{\overline{y}}

\newcommand{\pp}[2]{\frac{\partial #1}{\partial #2}}
\newcommand{\mS}{\mathcal{S}}
\newcommand{\ve}{\mathbf{e}}
\newcommand{\vv}{\mathbf{v}}
\newcommand{\vX}{\mathbf{X}}

\begin{document}

\newcommand{\arXivNumber}{2009.00670}

\renewcommand{\PaperNumber}{033}

\FirstPageHeading

\ShortArticleName{Invariants of Surfaces in Three-Dimensional Affine Geometry}

\ArticleName{Invariants of Surfaces\\ in Three-Dimensional Affine Geometry}

\Author{\"Orn ARNALDSSON~$^{\rm a}$ and Francis VALIQUETTE~$^{\rm b}$}

\AuthorNameForHeading{\"O.~Arnaldsson and F.~Valiquette}

\Address{$^{\rm a)}$~Department of Mathematics, University of Iceland, Reykjavik, Ssn. 600169-2039, Iceland}
\EmailD{\href{mailto:ornarnalds@hi.is}{ornarnalds@hi.is}}

\Address{$^{\rm b)}$~Department of Mathematics, Monmouth University, West Long Branch, NJ 07764, USA}
\EmailD{\href{mailto:fvalique@monmouth.edu}{fvalique@monmouth.edu}}

\ArticleDates{Received September 03, 2020, in final form March 21, 2021; Published online March 30, 2021}

\Abstract{Using the method of moving frames we analyze the algebra of differential invariants for surfaces in three-dimensional affine geometry. For elliptic, hyperbolic, and parabolic points, we show that if the algebra of differential invariants is non-trivial, then it is generically generated by a single invariant.}


\Classification{22F05; 53A35; 53A55}

\section{Introduction}

\looseness=1
The local geometry of $p$-dimensional submanifolds $S$ of an $m$-dimensional manifold $M$, under the smooth action of a Lie group $G$ is entirely governed by their differential invariants, in the sense that two submanifolds are locally congruent if and only if their differential invariants match~\cite{FO-1999,G-74}. A differential invariant is a (possibly locally defined) smooth function on the submanifold jet bundle $\J^{(\infty)}=\J^\infty(M,p)$ that remains unchanged under the prolonged action of~$G$. This prolonged action on $\J^{(\infty)}$ splits/reduces to an action on $G$-invariant subbundles (called \emph{branches of the equivalence problem}) whose symmetry properties differ; some branches having an infinite number of differential invariants of progressively higher and higher order while others have no invariants. The \emph{fundamental basis theorem}, first formulated in \cite[p.~760]{L-1893}, states that, on branches with non-trivial invariants, all the differential invariants can be generated from a finite number of low order invariants and their derivatives with respect to $p$ invariant total derivative operators $\D_1,\dots, \D_p$. For example, differential invariants of planar curves under the special Euclidean group ${\rm SE}(2)$ can all be expressed in terms of the curvature and its (repeated) arc-length derivatives~\cite{O-2001}. We~note that modern proofs of the fundamental basis theorem can be found in~\cite{KL-2006,KL-2016,OP-2007} and that this theorem is also frequently called the Lie--Tresse theorem.

A basic question, then, is to find a minimal generating set of invariants. According to the above, such a set will completely determine the local geometric properties of submanifolds under~$G$. The equivariant moving frame method is ideally suited for this type of question. Indeed, the effectiveness of the equivariant moving frame method lies in its recurrence relations, through which one obtains the complete and explicit structure of the underlying algebra of~differential invariants, and this without requiring explicit coordinate expressions for the moving frame or the invariants, leading to what is now referred to as the \emph{symbolic invariant calculus}~\cite{M-2010}. In~\cite{HO-2007} and~\cite{O-2009diff}, this was applied to deduce the surprising result that there is a single generating invariant for (suitably generic) surfaces in $\R^3$ under the projective, conformal, Euclidean and equi-affine groups. For the Euclidean and equi-affine groups, the algebra of differential invariants is generically governed by the Gaussian curvature and Pick invariant, respectively. Similarly, the algebra of differential invariants under the equi-affine group for generic parabolic surfaces with nonvanishing Pocchiola~4\textsuperscript{th} invariant has recently been shown to be generated by a single differential invariant in~\cite{CM-2020}.

In the current paper we study the geometry of surfaces under the entire affine group, ${\rm A}(3)={\rm GL}(3)\ltimes \R^3$, in detail. We~do not restrict ourselves to the most generic branch of surfaces as \mbox{in~\cite{HO-2007, O-2009diff}}, but rather provide all the different branches that have non-trivial invariants. In~each case, we study the algebra of differential invariants and obtain explicit formulas, in~terms of~surface jets, for the generating invariants and the invariants responsible for the various bran\-chings. In~certain cases, obtaining expressions for the invariants using the direct moving frame approach proved intractable. We~therefore relied on the recently developed technique of \emph{recursive moving frames}~\cite{O-2011}, to obtain the desired coordinate formulas. The main result of our paper is that whenever a branch admits differential invariants, the differential invariant algebra is~(gene\-rically) generated by a single invariant.

It is worth mentioning that, historically, differential geometers have given more attention to the problem of finding and classifying \emph{homogeneous spaces} within a given equivalence problem of submanifolds $S\subset M$ under the action of a Lie group $G$. We~recall that homogeneous spaces are, by definition, submanifolds that admit no non-trivial differential invariants, and, in a sense, the study of these spaces is the ``opposite" problem considered in this paper as we focus our attention to surfaces that admit non-trivial differential invariants. But for completeness, we note that the classification of homogeneous surfaces in $\mathbb{R}^3$ under the equi-affine group can be found in \cite[Theorem~12.4]{G-1963} and \cite[Chapter~VI]{J-1977}. More recently, normal forms for homogeneous surfaces in $\R^3$ under the general affine group with vanishing equi-affine Pick invariant were found in~\cite{ADV-1997}, and more generally in~\cite{DKR-1996} and~\cite{EE-1999}. We~note that since coordinate expressions for all relative and differential invariants derived in this paper are known, these could, theoretically, be used to find normal forms for the homogeneous surfaces. In~Section~\ref{sec:homogeneous surfaces} we provide several examples and show that a more efficient approach to deriving homogeneous surfaces is to integrate the moving frame equations. Though we emphasize that the study of homogeneous surfaces is not the main focus of the present paper.

We would be remiss if we failed to acknowledge the classical works of W.~Blaschke~\cite{B-1923}, and P.~Schirokov and A.~Schirokov~\cite{SS-1962} on the subject. Together with~\cite{DMMSV-1991}, and the references therein, they provide a classical treatment of affine differential geometry. The basic affine differential invariants can be found in these classical works, and the main contribution of our paper is the detailed analysis of the structure of the algebra of these differential invariants for surfaces in~affine 3-space.

We note that for parabolic surfaces, the problem studied in this paper is related to the local geometry of 2-nondegenerate real analytic hypersurfaces $S^5\subset \mathbb{C}^3$ in CR-geometry~\cite{M-2019}. This correspondence is not considered here, but we note that Question~7.1 in \cite[Section~7]{M-2019} is solved in this paper and corresponds to Case P.1.1 and its subcases. It is worth noting that~\cite{M-2019} has recently been superseded by the work of Doubrov, Merker, and The in~\cite{DMT-2020}.

\looseness=1 For a summary of the results obtained in this paper we refer the reader to Section \ref{sec:summary}. As for the rest of the paper, in Section \ref{sec:background} we recall the notion of a partial moving frame, introduce the recurrence relations that unlock the structure of the algebra of differential invariants, summarize the recursive moving frame implementation used to compute coordinate expressions of invariants, and finally recall basic results pertaining to the algebra of differential invariants. Sections \ref{sec:low-order normalizations}, \ref{sec:elliptic-hyperbolic}, and \ref{sec: parabolic} contain the main results of this paper. In Section~\ref{sec:low-order normalizations} we initiate the normalization process up to order two. At this order there is a splitting according to whether points are elliptic, hyperbolic, or parabolic. In~Section \ref{sec:elliptic-hyperbolic} we simultaneously consider elliptic and hyperbolic points. Finally, in Section \ref{sec: parabolic} we consider parabolic points.

\section{Background material}\label{sec:background}

In this section we recall basic results pertaining to the method of moving frames. We~refer the reader to the original manuscripts~\cite{FO-1999,KO-2003,O-2011} and the book~\cite{M-2010} for a more comprehensive exposition.

\subsection{Partial moving frames}\label{sec:partial moving frames}

In this section we introduce the notion of a partial moving frames as introduced in~\cite{O-2011}. Let~$G$ be an $r$-dimensional Lie group acting on an $m$-dimensional manifold $M$. We~are interested with the induced action of~$G$ on $p$-dimensional submanifolds $S\subset M$, where $1\leq p < m$ is fixed. For $0\leq n \leq \infty$, let $\J\n = \J\n(M,p)$ denote the $n$\textsuperscript{th} order submanifold jet bundle. Given the local coordinates $z=(x,u)=\big(x^1,\dots,x^p,u^1,\dots,u^q\big)$ on $M$, where $x$ are viewed as the independent variables and $u$ as the dependent variables, coordinates on $\J\n$ are given by $z\n = \big(x,u\n\big)=(\dots, x^i, \dots, u^\alpha_J, \dots)$, where $u^\alpha_J$ denote the derivative coordinates of orders $0\leq \#J\leq n$.

Let $\mS\n \subset \J\n$ be a $G$-invariant subbundle of $\J\n$ such that for all $g \in G$ near the identity, $g\cdot \mS\n \subseteq \mS\n$. Such an invariant subbundle is specified by a set of invariant differential equations
\begin{gather}\label{eq:submanifold}
\mS\n = \big\{ z\n \in \J\n\mid F\big(z\n\big)=0,\ \text{where}\ F\big(g\cdot z\n\big)\big|_{F(z\n)=0} = 0\big\}.
\end{gather}
The prolongation $\mS^{(n+1)}$ is obtained by appending the derivatives of the defining equations:
\begin{gather*}
\mS^{(n+1)} = \big\{z\n \in \J\n\mid F\big(z\n\big)=0,\, (D_1 F)\big(z^{(n+1)}\big)=0, \dots, (D_pF)\big(z^{(n+1)}\big)=0\big\},
\end{gather*}
where $D_i = D_{x^i}$ denote the total derivative operators. The induced action of $G$ on $\mS\n$ is called the \emph{$n$\textsuperscript{th} order prolonged action}. Borrowing Cartan's notational convention, we use capital letters to denote transformed variables: $Z\n = g\cdot z\n$. Let $\B\n = G\times \mS\n$ denote \emph{$n$\textsuperscript{th} order lifted bundle}. For $k\geq n$, we introduce the standard projection $\pi^k_n\colon \B^{(k)}\to \B\n$. The~lif\-ted bundle admits a groupoid structure with source map $\sigma\n\big(g,z\n\big)=z\n$ and target map $Z\n = \tau\n\big(g,z\n\big) = g\cdot z\n$ provided by the prolonged action. The action of $G$ on $\B\n$ is given by \emph{right-regularization}
\begin{gather*}
R_h\big(g,z\n\big) = \big(g\cdot h^{-1},h\cdot z\n\big).
\end{gather*}
Importantly, the target map $\tau\n\big(g,z\n\big)$ is \emph{invariant} under the right-regularized action. Therefore, the pull-back $\big(\tau\n\big)^*\eta$ of any differential form $\eta$ on $\mS\n$ is invariant on $\B\n$. Since the cotangent space $T^*\B\n=T^*G\times T^*\mS\n$ is a direct sum, and $G$ acts separately on its components, we may ``project'' any invariant 1-form on $\B\n$ to an invariant 1-form on $\mS\n$. Similarly, for higher order forms, we have the direct sums
\begin{gather*}
\bigwedge\nolimits^kT^*\B\n=\bigoplus_{i+j=k}\left(\bigwedge\nolimits^iT^*G \times \bigwedge\nolimits^jT^*\mS\n\right)\!,
\end{gather*}
which the right-regularized action preserves, and so we also have an invariant projection
\begin{gather*}
\pi_J\colon\ \bigwedge\nolimits^kT^*\B\n \to \bigwedge\nolimits^kT^*\mS\n
\end{gather*}
that maps invariant $k$-forms on $\B\n$ to invariant $k$-forms on $\mS\n$. In~practice we apply $\pi_J$ by~writing a $k$-form $\eta$ on $\B\n$ as a direct sum of wedge products of forms on $G$ and $\mS\n$ and then set all $T^*G$-terms (which in our case will be the Maurer--Cartan forms) to zero.

Given a differential form $\eta$ on $\mathcal{S}\n$, we introduce the \emph{lift map}
\begin{gather}\label{eq:lift}
\bl(\eta):=\pi_J\big(\tau\n\big)^*\eta,
\end{gather}
which returns an invariant form on $\B\n$ with only $T^*\mS\n$-components. The simplest example is given by the \emph{$n$\textsuperscript{th} order lifted invariants}
\begin{gather*}
\bl\big(z\n\big)=g\cdot z\n=Z\n.
\end{gather*}
\begin{Definition}
A \emph{partial right moving frame} of order $n$ is a right-invariant local subbundle $\widehat{\rho}\n \subset \B\n$, meaning that $R_h\big(\widehat{\rho}\n\big) \subset \widehat{\rho}\n$ for all $h\in G$.
\end{Definition}

In practice, a partial moving frame is obtained by choosing a cross-section $\K\n\subset \mS\n$ transversed to the prolonged group action. Then $\widehat{\rho}\n =\big(\tau\n\big)^{-1}(\mathcal{K}\n)$ is a partial moving frame of order $n$.

\begin{Remark}
We note that as opposed to the standard moving frame definition~\cite{FO-1999} a partial moving frame allows for some of the group parameters to not be normalized. More precisely, if~$\K\n\subset \mS\n$ has codimension $k_n$, then $\widehat{\rho}\n$ also has codimension~$k_n$, which implies that $r-k_n$ group parameters remain unnormalized.
\end{Remark}

Given a partial moving frame $\widehat{\rho}\n$, we introduce the \emph{partially normalized invariants}
\begin{gather*}
\widehat{Z}\n =\big(\widehat{\rho}\n\big)^*\big[\bl\big(z\n\big)\big].
\end{gather*}
The partially normalized invariants are obtained by substituting the normalized group parameters into the lifted invariants $Z\n$. To~simplify the notation in Sections \ref{sec:low-order normalizations}, \ref{sec:elliptic-hyperbolic}, and \ref{sec: parabolic}, we do not include the hat notation over the partially normalized invariants. We~hope that the context will make it clear that we are working with the partially normalized invariants.

\subsection{Recurrence relations}

The recurrence relations introduced in this section is one of the most important contributions of~\cite{FO-1999} to the method of moving frames. These equations unlock the structure of the algebra of~differential invariants (and more generally that of differential forms). One of the key aspects of~these equations is that they can be derived without the coordinate expressions for the (partial) moving frame, the differential invariants, and the invariant differential forms.

First, a coframe on $T^*\B^{(\infty)}$ is given by a basis of Maurer--Cartan forms $\mu^1,\dots,\mu^r$, the horizontal forms ${\rm d}x^1,\dots,{\rm d}x^p$, and the basic contact one-forms $\theta^\alpha_J = {\rm d}u^\alpha_J - u^\alpha_{J,j}{\rm d}x^j$. Throughout this paper we use the Einstein summation convention, where summation
occurs over repeated indices. Since all our computations are performed modulo contact forms, these are omitted from this point forward.

Applying the lift map~\eqref{eq:lift} to the horizontal coframe results in the invariant one-forms
\begin{gather*}
\omega^i = \bl\big({\rm d}x^i\big)
\end{gather*}
called \emph{lifted horizontal forms}.

Next, let
\begin{gather*}
\vv_\nu = \xi^i_\nu(z) \pp{}{x^i} + \phi^\alpha_\nu(z) \pp{}{u^\alpha},\qquad
\nu=1,\dots,r=\dim G,
\end{gather*}
be a basis of infinitesimal generators dual to the Maurer--Cartan form $\mu^1,\dots,\mu^r$. Then the recurrence relations for the lifted invariants measure the extend to which $d\circ \bl \neq \bl \circ d$. These equations are
\begin{gather}
{\rm d}X^i = \omega^i + \xi^i_\nu(Z) \mu^\nu,\nonumber
\\
{\rm d}U^\alpha_{J} = U_{J,j}^\alpha \omega^j + \phi^{\alpha;J}_\nu\big(Z^{(\#J)}\big) \mu^\nu,
\label{eq:lifted recurrence relations}
\end{gather}
where the prolonged vector field coefficients are given by the standard recursive formula
\begin{gather*}
\phi^{\alpha;J,j}_\nu = D_j \phi^{\alpha;J}_\nu - \big(D_j\xi^i_\nu\big)\cdot u^\alpha_{J,i}.
\end{gather*}

Given a partial moving frame $\widehat{\rho}\n$, which we can consider to be in $\B^{(\infty)}$ using the natural inclusion $i\n\colon \B\n \hookrightarrow \B^{(\infty)}$, we can then pull-back the lifted recurrence relations~\eqref{eq:lifted recurrence relations} by $\widehat{\rho}\n$ to obtain the recurrence relations for the partially normalized invariants
\begin{gather*}
{\rm d}\widehat{X}^i = \widehat{\omega}^i + \xi^i_\nu\big(\widehat{Z}\big) \widehat{\mu}^\nu,
\\
{\rm d}\widehat{U}^\alpha_{J} = \widehat{U}_{J,j}^\alpha \widehat{\omega}^j + \phi^{\alpha;J}_\nu\big(\widehat{Z}^{(\#J)}\big) \widehat{\mu}^\nu,
\end{gather*}
where
\begin{gather*}
\widehat{\omega}^i = \big(\widehat{\rho}\n\big)^*\omega^i\qquad\text{and}\qquad
\widehat{\mu}^\nu = \big(\widehat{\rho}\n\big)^*\mu^\nu
\end{gather*}
are the partially normalized horizontal one-forms and the partially normalized Maurer--Cartan forms, respectively.

\begin{Remark}
As in the standard moving frame implementation, the symbolic expressions for~the~par\-tially normalized Maurer--Cartan forms can be deduced from the recurrence relations for the phantom invariants, i.e., the lifted invariants that are equal to constant values by~vir\-tue of the moving frame construction. We~refer the reader to~\cite{FO-1999} for more detail.
\end{Remark}

\begin{Remark}
If the prolonged action becomes free on $\mS\n$, for a sufficiently large $n$, we note that the partial moving frame construction outlined above reproduces the usual moving frame construction first introduced in~\cite{FO-1999}. We~note that depending on~$\mS\n$, freeness cannot always be achieved and this even if the action is locally effective on subsets. Thus, Proposition~9.6 of~\cite{FO-1999} holds on regular subsets of the submanifold jet space but not necessarily on invariant subbundles of the form~\eqref{eq:submanifold}. When freeness cannot be attained, the most one can construct is a partial moving frame.
\end{Remark}

\subsection{Recursive moving frames}

For a detailed exposition of the recursive moving frame implementation, we refer the reader to the original work~\cite{O-2011}. One of the main issues of the standard moving frame implementation is that it first requires computing the prolonged action, which relies on implicit differentiation, and can lead to unwieldy expressions that limit the method's practical scope and implementation. This holds true even when using symbolic softwares such as {\sc Mathematica}, {\sc Maple}, or {\sc Sage}. Some of the results obtained in this paper are a prime example of this fact. Indeed, we implemented the standard moving frame machinery in {\sc Mathematica} and in some cases the software was unable to solve the normalization equations that produces the moving frame. In~those cases we had to revert to the recursive implementation.

The idea of the recursive moving frame method is, in the spirit of Cartan's original approach, to recursively normalize group parameters at a given order before prolonging the action to the next higher order jet space. Instead of using implicit differentiation to compute the prolonged action, the key idea of the recursive moving frame implementation is to use the recurrence formulas and the expressions for the Maurer--Cartan forms
\begin{gather}\label{eq:MC}
\bmu = {\rm d}g \cdot g^{-1}.
\end{gather}
To illustrate the recursive moving frame method, assume the prolonged action up to order $n$ is known and that a partial moving frame $\widehat{\rho}\n$ has been computed using a cross-section $\mathcal{K}\n \subset \mS\n$. Assuming, for simplicity, that $\mathcal{K}\n$ is a coordinate cross-section, suppose $u^\alpha_J=c$, with $\#J=n$ is one of the defining equation of $\mathcal{K}\n$. Then $\widehat{U}^\alpha_J=c$ is a phantom invariant and its recurrence relation yields
\begin{gather*}
0 = {\rm d}c = \widehat{U}^\alpha_{J,j} \widehat{\omega}^j + \phi^{\alpha;J}_\nu\big(\widehat{Z}\n\big)\widehat{\mu}^\nu
\end{gather*}
so that
\begin{gather}\label{eq:recursive prolongation}
\widehat{U}^\alpha_{J,j} \widehat{\omega}^j = - \phi^{\alpha;J}_\nu\big(\widehat{Z}\n\big)\widehat{\mu}^\nu.
\end{gather}
By assumption, coordinate expressions for $\phi^{\alpha;J}_\nu(\widehat{Z}\n)$ are known, since the prolonged action up to order $n$ has been computed, and the partially normalized Maurer--Cartan forms $\widehat{\mu}^\nu$ can be found by substituting the group normalizations into~\eqref{eq:MC}. Expressing the right-hand side of~\eqref{eq:recursive prolongation} as a linear combination of the partially normalized horizontal forms $\widehat{\omega}^i$, we are able to obtain expressions for the order $n+1$ partially normalized invariants $\widehat{U}^\alpha_{J,j}$.

\subsection{The algebra of differential invariants}

Assume a moving frame is known or that a partial moving frame has been computed with no possibility of further group parameter normalizations. Dual to the invariant horizontal forms $\omega^i$ are the \emph{invariant total derivative operators}
\begin{gather}\label{eq:invariant derivative operators}
\D_i = \widehat{W}^j_i D_i,\qquad \text{where}\quad \big(\widehat{W}^j_j\big) = \big(\widehat{\rho}\n\big)^*\big(D_j X^i\big)^{-1}.
\end{gather}
Now, let
\begin{gather}\label{eq:omega eq}
{\rm d}\omega^i = C^i_{jk} \omega^j\wedge\omega^k
\mod(\text{unnormalized Maurer--Cartan forms})
\end{gather}
be the structure equations among the invariant horizontal forms. These equations can be obtai\-ned symbolically by extending the recurrence relations~\eqref{eq:lifted recurrence relations} to differential forms as done in~\cite{KO-2003}. Given~\eqref{eq:omega eq}, the commutation relations among the invariant total derivative opera\-tors~are
\begin{gather}\label{eq:commutator}
[\D_j,\D_k]=-C^i_{jk} \D_i.
\end{gather}
Fix $j$, $k$ in~\eqref{eq:commutator} and apply the commutation relation to $p$ invariants $I_1,\dots,I_p$
to obtain $[\D_j,\D_k]I_\ell = -C^i_{jk} \D_i I_\ell$. In~matrix form
\begin{gather*}
[\D_j,\D_k] I = - \D I C_{jk},
\end{gather*}
where $[\D_j,\D_k] I = ([\D_j,\D_k] I_1,\dots,[\D_j,\D_k] I_p)^{\rm T}$, $\D I = (\D_i I_\ell)$, and $C_{jk} = \big(C^1_{jk},\dots, C^p_{jk}\big)^{\rm T}$. If~$\det \D I \not\equiv 0$, then one can solve for $C_{ij}$
\begin{gather}\label{eq:commutator trick}
C_{jk} = -(\D I)^{-1} [\D_j,\D_k] I,
\end{gather}
which allows one to express the commutator invariants $C_{jk}$ in terms of $I=(I_1,\dots,I_p)$ and its invariant derivatives. This is what we refer to as the \emph{commutator trick}. Notice that given a~single invariant $I_1$, we could have set $I_i:=\D_{k_i}^{\ell_i} I_1$, with $1\leq k_i \leq p$ and $\ell_i\geq 0$, in order to write the commutator invariants $C_{jk}$ as functions of a single invariant and its invariant derivatives. This observation plays a key role in showing that the algebras of differential invariants for Euclidean, equi-affine, conformal, and projective surfaces are generically generated by a single invariant~\cite{HO-2007,O-2009diff,O-2016}. The commutator trick will also be used in this paper to show that certain algebras of differential invariants are generated by a single invariant.

We now recall important results about the algebra of differential invariants that can be found in~\cite{FO-1999,O-2007}.

\begin{Proposition}\label{prop:complete set}
The normalized invariants $\widehat{Z}\n$ provide a complete set of differential invariants of order $\leq n$.
\end{Proposition}

By the \emph{replacement principle}~\cite{FO-1999,M-2010}, if $I\big(z\n\big)$ is a differential invariant, then it can be written in terms of the normalized invariants as $I=I\big(\widehat{Z}\n\big)$, which is obtained by replacing the jet coordinates $z\n$ by their corresponding normalized invariants $\widehat{Z}\n$.

\begin{Definition}
A set of invariants $\bm{I}_{{\rm gen}}=\{I_1,\dots,I_\ell\}$ is said to \emph{generate} the algebra of dif\-fe\-ren\-tial invariants if any differential invariant can be expressed in terms of $\bm{I}_{{\rm gen}}$ and its invariant derivatives~\eqref{eq:invariant derivative operators} of any order.
\end{Definition}

From Proposition \ref{prop:complete set} it follows that if one can show that the normalized invariants $\widehat{Z}^{(\infty)}$ can be written in terms of a set of invariants $\bm{I}_{{\rm gen}}$ and its invariant derivatives, then $\bm{I}_{{\rm gen}}$ is a~generating set for the algebra of differential invariants.

\begin{Theorem}\label{thm:generating set}
Given a moving frame $\widehat{\rho}\n$, the normalized invariants $\bm{I}_{{\rm gen}} = \big\{\widehat{Z}^{(n+1)}\big\}$ form a~generating set of differential invariants.
\end{Theorem}

The generating set in Theorem \ref{thm:generating set} is not necessarily minimal. By that we mean that it might be possible to remove certain non-phantom invariants and still obtain a generating set. To~this day, there is no known result that stipulates how small the generating set can be. But if one can show that the invariants $\bm{I}_{{\rm gen}}=\big\{\widehat{Z}^{(n+1)}\big\}$ can be expressed in terms of a single invariant $I$ and its invariant derivatives $\D_1,\dots,\D_p$, then the algebra of differential invariants is generated by a single function. This is the approach used in the following sections to show that the various differential invariant algebras are generated by a single invariant.

\section{Affine action and low-order normalizations}\label{sec:low-order normalizations}

In the following, we consider surfaces $S\subset \mathbb{R}^3$, which we assume are locally given a graphs of~functions:
\begin{gather*}
S = \{z=(x,y,u(x,y))\} \subset \mathbb{R}^3.
\end{gather*}
We are interested in the action of the affine group ${\rm A}(3,\mathbb{R}) = {\rm GL}(3,\mathbb{R})\ltimes \mathbb{R}^3$ on these surfaces given by
\begin{gather*}
Z = Az+b,\qquad\text{where}\quad A\in {\rm GL}(3,\mathbb{R})\quad\text{and}\quad b\in \mathbb{R}^3.
\end{gather*}
A basis for the algebra of infinitesimal generators is provided by
\begin{gather*}
\vv_{xx} = x\pp{}{x},\quad\ \vv_{xy} = y\pp{}{x},\quad\ \vv_{xu} = u \pp{}{x},\quad\ \vv_{yx} = x\pp{}{y},\quad\ \vv_{yy} = y\pp{}{y},\quad\ \vv_{yu} = u\pp{}{y},
\\
\vv_{ux} = x\pp{}{u},\quad\ \vv_{uy} = y\pp{}{u},\quad\ \vv_{uu} = u\pp{}{u},\quad\ \vv_x = \pp{}{x},\quad\ \vv_y = \pp{}{y},\quad\ \vv_u = \pp{}{u}.
\end{gather*}
Let
\begin{gather*}
\bmu=\begin{bmatrix}
\mu &\nu \\
0 & 0
\end{bmatrix}\qquad\text{with}\quad
\mu = \begin{bmatrix}
\mu^{xx} & \mu^{xy} & \mu^{xu} \\
\mu^{yx} & \mu^{yy} & \mu^{yu} \\
\mu^{ux} & \mu^{uy} & \mu^{uu}
\end{bmatrix}
\quad\text{and}\quad
\nu = \begin{bmatrix}
\mu^x \\ \mu^y \\ \mu^u
\end{bmatrix}
\end{gather*}
denote a basis of Maurer--Cartan forms with structure equations
\begin{gather*}
{\rm d}\mu = -\mu \wedge \mu,\qquad
{\rm d}\nu = - \mu\wedge \nu.
\end{gather*}
Then the order zero recurrence relations for the lifted invariants are
\begin{gather*}
{\rm d}X = \omega^x + X \mu^{xx} + Y\mu^{xy} + U \mu^{xu} + \mu^x,
\\
{\rm d}Y = \omega^y + X \mu^{yx} + Y\mu^{yy} + U\mu^{yu} + \mu^y,
\\
{\rm d}U = U_j \omega^j + X\mu^{ux} + Y \mu^{uy} + U\mu^{uu} + \mu^u,
\end{gather*}
while for $k+\ell\geq 1$,
\begin{gather*}
{\rm d}U_{X^kY^\ell} = U_{X^kY^\ell j} \omega^j - k U_{X^k Y^\ell} \mu^{xx} - \ell U_{X^{k+1}Y^{\ell-1}} \mu^{xy} - k U_{X^{k-1}Y^{\ell+1}} \mu^{yx} - \ell U_{X^k Y^\ell} \mu^{yy}
\\ \hphantom{{\rm d}U_{X^kY^\ell} =}
{} + U_{X^k Y^\ell} \mu^{uu} + \delta_{1k}\delta_{0\ell} \mu^{ux} + \delta_{0k}\delta_{1\ell} \mu^{uy}
\\ \hphantom{{\rm d}U_{X^kY^\ell} =}
{} - \sum_{\substack{0\leq i \leq k\\0\leq j\leq \ell\\ (i,j)\neq (k,\ell)}} \binom{k}{i}\binom{\ell}{j} \big[ U_{X^{k-i}Y^{\ell-j}} U_{X^{i+1}Y^j} \mu^{xu} + U_{X^{k-i}Y^{\ell-j}} U_{X^i Y^{j+1}} \mu^{yu}\big],
\end{gather*}
where there is no summation over $k$ and $\ell$, and $\delta_{ij}$ denotes the Kronecker delta function.

Since the action is transitive on $\J^{(1)}$, we can set
\begin{gather}\label{eq: initial normalizations}
X = Y = U = U_X = U_Y = 0.
\end{gather}
In other words, we can choose the cross-section $\mathcal{K}^{(1)}=\{x=y=u=u_x=u_y=0\}\subset \J^{(1)}$. The~recurrence relations for these phantom invariants are
\begin{gather*}
0 = \omega^x + \mu^x,\qquad
0 = \omega^y + \mu^y,\qquad
0 = \mu^u,\qquad
0 = U_{Xj} \omega^j + \mu^{ux},\qquad
0 = U_{Yj} \omega^j + \mu^{uy}.
\end{gather*}
As mentioned in Section \ref{sec:partial moving frames}, from this point onward we omit the use of the hat notation to denote partially normalized quantities. Solving for the Maurer--Cartan forms yields
\begin{gather}\label{eq: initial mc normalizations}
\mu^x = -\omega^x,\qquad \mu^y = -\omega^y,\qquad \mu^u = 0,\qquad
\mu^{ux} = -U_{Xj} \omega^j,\qquad \mu^{uy} = -U_{Yj} \omega^j.
\end{gather}
Taking into account the order 0 and 1 normalizations~\eqref{eq: initial normalizations}, and the normalized Maurer--Cartan forms~\eqref{eq: initial mc normalizations}, the recurrence relations for the order 2 partially normalized invariants are
\begin{gather}
{\rm d}U_{XX} = U_{XXj} \omega^j + U_{XX}\big(\mu^{uu} - 2\mu^{xx}\big) - 2U_{XY} \mu^{yx},\nonumber
\\
{\rm d}U_{XY} = U_{XYj} \omega^j - U_{XX}\mu^{xy} + U_{XY}\big(\mu^{uu} - \mu^{xx} - \mu^{yy}\big) - U_{YY}\mu^{yx},\nonumber
\\
{\rm d}U_{YY} = U_{YYj} \omega^j + U_{YY}(\mu^{uu}-2\mu^{yy}) - 2U_{XY}\mu^{xy}.
\label{eq: order 2 recurrence relations}
\end{gather}
Consider the partially normalized lifted Hessian determinant
\begin{gather*}
H=U_{XX}U_{YY} - U_{XY}^2.
\end{gather*}
Since
\begin{gather*}
{\rm d}H = 2 H\big(\mu^{uu} - \mu^{xx} - \mu^{yy}\big) \mod(\omega^x,\omega^y),
\end{gather*}
we conclude that $H$ is a relative invariant. To~obtain an expression for $H$, we introduce the determinant
\begin{gather}\label{eq:det}
|D\vX| = \det
\begin{bmatrix}
X_x & X_y \\
Y_x & Y_y
\end{bmatrix} = \det
\begin{bmatrix}
a_{11} + a_{13} u_x & a_{12} + a_{13} u_y \\
a_{21} + a_{23} u_y & a_{22} + a_{23} u_y
\end{bmatrix}
\end{gather}
and the Hessian determinant $h=u_{xx} u_{yy} - u_{xy}^2$. Then
\begin{gather*}
H = \frac{a_{33}^2}{|D\vX|^2} h.
\end{gather*}

\begin{Definition}
A point $\big(x,y,u^{(2)}\big)$ of $S^{(2)} \in \J^{(2)}$ is said to be
\begin{itemize}
\item \emph{elliptic} if $h>0$,
\item \emph{hyperbolic} if $h<0$,
\item \emph{parabolic} if $h=0$.
\end{itemize}
\end{Definition}

The remaining analysis depends on the sign of the Hessian determinant. Since most results for elliptic and hyperbolic points are similar, these two cases are combined together in the next section. The case of parabolic points is considered in Section \ref{sec: parabolic}.

\section{Elliptic and hyperbolic points}\label{sec:elliptic-hyperbolic}

In this section we work under the assumption that
\begin{gather*}
H = \epsilon = \pm 1,
\end{gather*}
with $\epsilon=1$ corresponding to the elliptic case and $\epsilon = -1$ to hyperbolic points. From the recurrence relations~\eqref{eq: order 2 recurrence relations}, we conclude that it is possible to set
\begin{gather}\label{eq: EH order 2 normalizations}
U_{XX} = 1,\qquad U_{YY} = \epsilon,\qquad U_{XY}=0.
\end{gather}

\begin{Remark}\label{remark:ambiguity}
In Cartesian coordinates, the normalization equations~\eqref{eq: EH order 2 normalizations} are quadratic in the group parameters. Therefore, in the process of constructing a moving frame there is a choice of sign that needs to be made. But since~\eqref{eq: EH order 2 normalizations} holds, no matter the choice made, this does not affect the algebra of differential invariants of the surface and as such is not important for our purpose. Thus, as it is customary~\cite{O-2016}, in the following we omit such ambiguity.
\end{Remark}

After the normalizations~\eqref{eq: EH order 2 normalizations} have been performed, the recurrence relations for the order 3 partially normalized invariants
are
\begin{gather*}
{\rm d}U_{X^3}= -3 \mu^{xu} - \frac{U_{X^3}}{2}\mu^{uu} \mod (\omega^x,\omega^y),
\\
{\rm d}U_{X^2Y}= -\epsilon \mu^{yu}+\epsilon U_{X^3} \mu^{yx} -2U_{XY^2}\mu^{yx}-\frac{U_{X^2Y}}{2}\mu^{uu}\mod (\omega^x,\omega^y),
\\
{\rm d}U_{XY^2}= -\epsilon \mu^{xu} + 2\epsilon U_{X^2Y}\mu^{yx}-U_{Y^3}\mu^{yx} -\frac{U_{XY^2}}{2}\mu^{uu}\mod (\omega^x,\omega^y),
\\
{\rm d}U_{Y^3}= -3\mu^{yu}+3\epsilon U_{XY^2}\mu^{yx}-\frac{U_{Y^3}}{2}\mu^{uu}\mod (\omega^x,\omega^y).
\end{gather*}
Consistent with normalizations performed for elliptic and hyperbolic surfaces in equi-affine geo\-metry~\cite{O-2009diff}, we set
\begin{gather*}
U_{X^3}+\epsilon U_{XY^2} = U_{Y^3}+\epsilon U_{X^2Y}=0
\end{gather*}
and solve for $U_{XY^2}$ and $U_{X^2Y}$. We~are then left with $U_{X^3}$ and $U_{Y^3}$, whose recurrence rela\-tions~are
\begin{gather*}
{\rm d}U_{X^3} =3\epsilon U_{Y^3}\mu^{yx} - \frac{U_{X^3}}{2}\mu^{uu} \mod(\omega^x,\omega^y),
\\
{\rm d}U_{Y^3} = -3U_{X^3}\mu^{yx} - \frac{U_{Y^3}}{2}\mu^{uu} \mod(\omega^x,\omega^y).
\end{gather*}
The extent to which one can solve for the partially normalized Maurer--Cartan forms $\mu^{yx}$ and~$\mu^{uu}$ depends on the determinant
\begin{gather*}
\det \begin{bmatrix}
3\epsilon U_{Y^3} & - \dfrac{U_{X^3}}{2} \\[1.5ex]
-3U_{X^3} & - \dfrac{U_{Y^3}}{2}
\end{bmatrix} = -\frac{3}{2} \big(U_{X^3}^2+ \epsilon U_{Y^3}^2\big) = - \frac{3}{2} P_\epsilon.
\end{gather*}
We note that $P_\epsilon$ is a relative invariant as
\begin{gather*}
{\rm d}P_\epsilon = -P_\epsilon \mu^{uu}.
\end{gather*}
In fact, $P_\epsilon=\frac{P}{a_{33}}$, where $P$ is the equi-affine Pick invariant
\begin{gather*}
P =\frac{1}{16 \big(u_{xx} u_{yy}-u_{xy}^2\big)^3}\big[6 u_{xx}u_{xy} u_{yy} u_{xxx}u_{yyy}-6_{xx}u_{yy}^2 u_{xxx} u_{xyy} -18 u_{xx}u_{xy} u_{yy} u_{xxy} u_{xyy}
\\ \hphantom{P =}
{}+12 u_{xx}u_{xy}^2 u_{xxy} u_{yyy}-6 u_{xx}^2u_{yy} u_{xxy}u_{yyy}+9 u_{xx}u_{yy}^2 u_{xxy}^2 -6
 u_{xx}^2 u_{xy} u_{xyy} u_{yyy}
\\ \hphantom{P =}
{}+9 u_{xx}^2 u_{yy} u_{xyy}^2 +u_{xx}^3 u_{yyy}^2-6u_{xy} u_{yy}^2 u_{xxx} u_{xxy} +12 u_{xy}^2 u_{yy} u_{xxx} u_{xyy}
\\ \hphantom{P =}
{}-8 u_{xy}^3 u_{xxx} u_{yyy} + u_{yy}^3 u_{xxx}^2 \big].
\end{gather*}

We now need to distinguish the cases where $P_{\epsilon}\equiv 0$ is identically zero and where $P_{\epsilon}\neq 0$ does not vanish. In~the elliptic case, we note that if $P_1 \equiv 0$, then $U_{X^3}\equiv U_{Y^3}\equiv 0$. On the other hand, in the hyperbolic case, when $P_{-1} \equiv 0$, we have that $U_{Y^3} \equiv \pm U_{X^3}$. But, we observe that under the change of variables $(x,y,u)\mapsto (x,-y,u)$, we can always assume that $U_{Y^3}=-U_{X^3}$. Therefore, at hyperbolic points there are two cases to consider, either $U_{X^3} \equiv 0$ or $U_{X^3} \neq 0$. We~combine the different cases as follows:
\begin{gather*}
{\bf EH.1}\colon\ P_\epsilon \neq 0,\qquad
{\bf EH.2}\colon\ U_{X^3}\equiv U_{Y^3} \equiv 0,\qquad
{\bf H.3}\colon\ U_{Y^3} \equiv - U_{X^3}\neq 0.
\end{gather*}
We note that cases EH.1 and EH.2 hold for both elliptic and hyperbolic points whereas case H.3 is only for hyperbolic points. In~local coordinates, since
\begin{gather*}
U_{XXX} = \frac{C_1\big(3\epsilon a_{33} u_{xx}Y_x - 4Y_x^3\big)-C_2\big(\epsilon a_{33}u_{xx}-4 Y_x^2\big)\sqrt{|h|}\sqrt{a_{33}u_{xx} - \epsilon Y_x^2}}{4a_{33}^2u_{xx}^3 |h|^{3/2}},
\\[1ex]
U_{YYY} = \frac{C_1 \big(a_{33}u_{xx}-4\epsilon Y_x^2\big)\sqrt{a_{33}u_{xx}-\epsilon Y_x^2}+C_2\sqrt{|h|}\big(3\epsilon a_{33}u_{xx}Y_x - 4 Y_x^3\big)}{4a_{33}^2u_{xx}^3|h|^{3/2}},
\end{gather*}
and
\begin{gather*}
P_\epsilon = \frac{C_1^2 + hC_2^2}{16 a_{33}u_{xx}^3h^3},
\end{gather*}
where
\begin{gather*}
C_1 = 6u_{xx}u_{xy}^2u_{xxy} - 4u_{xy}^3 u_{xxx} - 3u_{xx}^2u_{xy}u_{xyy}-3u_{xx}^2u_{yy}u_{xxy}+3u_{xx}u_{xy}u_{yy}u_{xxx}+u_{xx}^3u_{yyy},
\\
C_2 = -6u_{xx}u_{xy}u_{xxy}+4u_{xy}^2u_{xxx}+3u_{xx}^2u_{xyy}-u_{xx}u_{yy}u_{xxx}.
\end{gather*}
the three cases can be restated as
\begin{gather*}
{\bf EH.1}\colon\ C_1^2 + hC_2^2\neq 0,\qquad
{\bf EH.2}\colon\ C_1 \equiv C_2 \equiv 0,\qquad
{\bf H.3}\colon\ C_1 \equiv - C_2 \sqrt{|h|}\neq 0.
\end{gather*}

\begin{Remark}
We remark that the expressions for $U_{XXX}$ and $U_{YYY}$ hold provided $u_{xx}\neq 0$. From this point forward, we always work on the open dense subset of the jet space where $u_{xx}\neq 0$.
\end{Remark}

\subsection{Case EH.1}

When $P_\epsilon \neq 0$, it is possible to set
\begin{gather*}
U_{X^3} =1,\qquad U_{Y^3}=0.
\end{gather*}
According to Theorem \ref{thm:generating set}, the order 4 differential invariants
\begin{gather*}
U_{X^4}, \qquad U_{X^3Y}, \qquad U_{X^2Y^2}, \qquad U_{XY^3}, \qquad U_{Y^4},
\end{gather*}
form a complete set of generating invariants. We~now show in fact that the algebra of differential invariants is generically generated by the single invariant $I_1= U_{Y^4}$. First, the structure equations for the invariant coframe $\omega^x$, $\omega^y$ are
\begin{gather*}
{\rm d}\omega^x = \frac{2\epsilon}{3} U_{XY^3} \omega^x\wedge\omega^y,\qquad
{\rm d}\omega^y = \frac{1}{12}\big(3U_{X^4}-6\epsilon U_{X^2Y^2} - U_{Y^4}\big) \omega^x\wedge\omega^y.
\end{gather*}
Therefore, the Lie bracket of the invariant total derivative operators is
\begin{gather*}
[\D_x,\D_y] = -\frac{2\epsilon}{3} U_{XY^3} \D_x - \frac{1}{12}\big(3U_{X^4}-6\epsilon U_{X^2Y^2} - U_{Y^4}\big) \D_y.
\end{gather*}
Using the commutator trick~\eqref{eq:commutator trick}, we can generically solve for $I_2=U_{XY^3}$ and $I_3=U_{X^4}-2\epsilon U_{X^2Y^2}$ in terms of $I_1$ and its invariant derivatives. Indeed, applying the commutator trick to $I_1$ and~$\D I_1$, where $\D$ is a nontrivial invariant total derivative operator, we find that
\begin{gather*}
\begin{pmatrix}
-\frac{2\epsilon I_2}{3} \\[.5ex]
\frac{I_1}{12}-\frac{I_3}{4}
\end{pmatrix}
=
\begin{pmatrix}
\D_x I_1 & \D_y I_1 \\[.5ex]
\D_x\D I_1 & \D_y\D I_1
\end{pmatrix}^{-1}
\begin{pmatrix}
[\D_x,\D_y]I_1 \\[.5ex]
[\D_x,\D_y]\D I_1
\end{pmatrix},
\end{gather*}
which can be solved for $I_2$ and $I_3$ provided that
\begin{gather*}
\D_xI_1\cdot \D_y\D I_1 - \D_yI_1\cdot \D_x\D I_1\neq 0.
\end{gather*}
Next, consider the syzygy
\begin{gather}\label{eq:EH syzygy}
\mathcal{D}_xI_1-\mathcal{D}_yI_2 = \frac{3}{2}I_3 - \frac{7\epsilon}{6}I_2^2 - \frac{1}{2}I_1I_3+\frac{1}{2}I_1^2+\frac{1}{4}\big(3U_{X^2Y^2}^2 + 6\epsilon U_{X^2Y^2}-2U_{X^3Y}U_{XY^3}\big).
\end{gather}
This suggests the introduction of the fourth order invariant
\begin{gather*}
I_4=3U_{X^2Y^2}^2+6U_{X^2Y^2} - 2U_{X^3Y}U_{XY^3}.
\end{gather*}
Also, from~\eqref{eq:EH syzygy} is follows that $I_4$ can be expressed in terms of $I_1$, $I_2$, $I_3$ and their invariant derivatives. Since $I_2$ and $I_3$ can be expressed in terms of $I_1$ and its invariant derivatives, the same holds true for $I_4$.

Now, considering the fifth order invariants $\D_i I_j$, we find, using {\sc Mathematica}, the syzygy
\begin{gather*}
-216 I_2 \D_xI_2 - 108\epsilon I_2\D_yI_3 + 36\epsilon I_2 \D_yI_4 + 216I_2^2 - 36I_2^2\D_xI_3 + 12I_2^2\D_xI_4 + 54 I_1I_2^2
\\ \qquad
{}+ 48\epsilon I_2^3\D_xI_2 + 24 I_2^3\D_yI_3 + 36\epsilon I_2^4 - 4\epsilon I_1I_2^4 - 108I_2^2I_3 + 6I_1I_2^2I_3-10\epsilon I_2^4I_3
\\ \qquad
{}-36\epsilon I_4\D_yI_2 - 9\epsilon I_1I_4 - 12I_2I_4\D_xI_2 - 30I_2^2I_4 - 2I_1I_2^2I_4 + 3I_2^2I_3I_4
\\ \qquad
{}+ \big(216\epsilon \D_yI_2+54\epsilon I_1+ 72I_2\D_xI_2
- 432\epsilon I_2\D_xI_2 - 108I_2\D_yI_3 + 36I_2\D_yI_4
\\ \qquad\qquad
{}+ 180I_2^2 + 270\epsilon I_2^2 + 12 I_1I_2^2+ 66\epsilon I_1I_2^2 - 2I_2^4 -18I_2^2I_3-198\epsilon I_2^2I_3+6\epsilon I_1I_2^2I_3
\\ \qquad\qquad
{}-27I_4 - 36I_4\D_yI_2 - 18I_1I_4 -33\epsilon I_2^2I_4\big)U_{X^2Y^2}
\\ \qquad
{}+ \big(162 + 2166 \D_yI_2 + 108\epsilon \D_yI_2 + 108I_1+27\epsilon I_1
-180I_2\D_xI_2-144I_2^2+198\epsilon I_2^2
\\ \qquad\qquad
{}+18I_1I_2^2 -99I_2^2I_3-54\epsilon I_4-9\epsilon I_1I_4\big)U_{X^2Y^2}^2
\\ \qquad
{}+ \big(81+324\epsilon +108\D_yI_2 + 54I_1+54\epsilon I_1 -189\epsilon I_2^2
- 27I_4\big)U_{X^2Y^2}^3
\\ \qquad
{}+ (162+162\epsilon+27\epsilon I_1)U_{X^2Y^2}^4 + 81U_{X^2Y^2}^5 = 0.
\end{gather*}
This is a quintic equation in $U_{X^2Y^2}$, which can locally be solved in terms of $I_1$, $I_2$, $I_3$, and $I_4$ and their invariant derivatives. This shows the following results.

\begin{Theorem}
If the equi-affine Pick invariant $P\neq0$ does not vanish, then the algebra of~dif\-ferential invariants is generically generated by the fourth order invariant $I_1=U_{Y^4}$.
\end{Theorem}

Using the method of recursive moving frames, a coordinate expression for the generating invariant is
\begin{gather*}
U_{Y^4} = 3 + \frac{3}{P\sqrt{|h|}}(L \D_y K - K \D_yL) +\frac{3}{4P}\D_y\big(P \D_y(\ln|h|)\big)+ \frac{3\epsilon}{16}\big(\D_y(\ln|h|)\big)^2
\\ \hphantom{U_{Y^4} =}
{}+\frac{3\D_y(\ln|h|)}{4P\sqrt{|h|}}(J \D_y K-I \D_y L) + \frac{3\epsilon\D_x(\ln|h|)}{4P\sqrt{|h|}}(K \D_y L - L \D_yK),
\end{gather*}
where
\begin{gather*}
\D_x = \frac{1}{P\sqrt{|h|}}(L D_x - K D_y),\qquad
\D_y = \frac{1}{P\sqrt{|h|}}(-J D_x + I D_y)
\end{gather*}
are invariant total derivative operators and
\begin{gather}
I = \sqrt{Pu_{xx}-\epsilon K^2},\qquad
J = \frac{u_{xy}\sqrt{Pu_{xx}-\epsilon K^2}-\epsilon \sqrt{|h|} K}{u_{xx}},\nonumber
\\
L = \frac{u_{xy}K+\sqrt{|h|}\sqrt{Pu_{xx}-\epsilon K^2}}{u_{xx}},
\label{eq:IJL}
\end{gather}
with $K$ a solution to the sextic equation
\begin{gather}\label{eq:bi-cubic}
16\epsilon K^6 - 24 (Pu_{xx})K^4+ 9\epsilon (Pu_{xx})^2 K^2 - \frac{(Pu_{xx})^3 C_1^2}{C_1^2+hC_2^2} =0.
\end{gather}

\begin{Remark}
Over the real numbers, the bi-cubic equation~\eqref{eq:bi-cubic} has one real solution for~$K^2$. Then, as in Remark \ref{remark:ambiguity} there is an ambiguity of sign in the definition of $K$, but this does not affect the structure of the algebra of differential invariants. Also, on the cross-section, equation~\eqref{eq:bi-cubic} reduces to
\begin{gather}\label{eq:reduced bi-cubic}
16\epsilon K^6-24K^4+9\epsilon K^2 = 0
\end{gather}
so that $K^2=0,\frac{3}{4\epsilon}$. Perturbing~\eqref{eq:reduced bi-cubic} near the cross-section, the zero root becomes positive, which implies that $K$ is defined near the cross-section. Finally, we note that on the cross-section $Pu_{xx}-\epsilon K^2 = 1$ so that the square roots occurring in~\eqref{eq:IJL} are well-defined in the neighborhood of the cross-section.
\end{Remark}

\subsection{Case EH.2}

We are now assuming that $U_{X^3} \equiv U_{Y^3}\equiv 0$. Their recurrence relations imply that
\begin{gather}\label{eq:EH.2 constraints}
U_{X^4} \equiv 3\epsilon U_{X^2Y^2}\equiv U_{Y^4},\qquad U_{X^3Y}\equiv U_{XY^3}\equiv 0.
\end{gather}
Thus, there is only one fourth order partially normalized invariant. We~continue the analysis using the invariant
\begin{gather*}
U_{X^2Y^2} = \frac{18u_{xx}u_{xy}u_{xxx}u_{xxy}\!-\!9u_{xx}^2u_{xxy}^2\!-\!(4u_{xy}^2 \!+\!5u_{xx}u_{yy})u_{xxx}^2\!+\!3u_{xx}(u_{xx}u_{yy}\!-\!u_{xy}^2)u_{xxxx}}{9 a_{33} u_{xx}^3(u_{xx}u_{yy}-u_{xy}^2)}.
\end{gather*}
Since its recurrence relation is
\begin{gather*}
{\rm d}U_{X^2Y^2} = - U_{X^2Y^2}\mu^{uu} \mod(\omega^x,\omega^y),
\end{gather*}
we now have to consider the cases
\begin{gather*}
{\bf EH.2.1}\colon\ U_{X^2Y^2}\neq 0,\qquad
{\bf EH.2.2}\colon\ U_{X^2Y^2}\equiv 0.
\end{gather*}

\subsubsection{Case EH.2.1}

When $U_{X^2Y^2}\neq 0$, we can normalize
\begin{gather*}
U_{X^2Y^2} =1.
\end{gather*}
From~\eqref{eq:EH.2 constraints} is follows that all fourth order invariants are constant
\begin{gather*}
U_{X^4}\equiv U_{Y^4} \equiv 3\epsilon,\qquad U_{X^2Y^2}=1,\qquad U_{X^3Y}\equiv U_{XY^3}\equiv 0.
\end{gather*}
Considering their recurrence relations
\begin{gather*}
0 \equiv {\rm d}U_{X^4} = \big(U_{X^5}-3\epsilon U_{X^3Y^2}\big)\omega^x+\big(U_{X^4Y}-3\epsilon U_{X^2Y^3}\big)\omega^y,\\
0 \equiv {\rm d}U_{X^3Y} = U_{X^4Y}\omega^x + U_{X^3Y^2}\omega^y,\\
0 \equiv {\rm d}U_{XY^3} = U_{X^2Y^3}\omega^x + U_{XY^4}\omega^y,\\
0 \equiv {\rm d}U_{Y^4} = (U_{XY^4} - 3\epsilon U_{X^3Y^2})\omega^x + \big(U_{Y^5}-3\epsilon U_{X^2Y^3}\big)\omega^y,
\end{gather*}
we find that all fifth order invariants vanish. Similarly, the recurrence relations for the fifth order invariants imply that the sixth order invariants are constant, and so on. Therefore, all the invariants are constant and there are no further normalizations possible. In~particular, the Maurer--Cartan form $\mu^{yx}$ cannot be normalized. The structure equations for the coframe $\{\omega^x, \omega^y, \mu^{yx}\}$ are
\begin{gather*}
{\rm d}\omega^x = -\epsilon \mu^{yx}\wedge \omega^y,\qquad
{\rm d}\omega^y = \mu^{yx}\wedge \omega^x,\qquad
{\rm d}\mu = \epsilon \omega^y \wedge \omega^x.
\end{gather*}

\subsubsection{Case EH.2.2}

When $U_{X^2Y^2}\equiv 0$, the same argument as in Case EH.2.1 implies that all higher order partially normalized invariants vanish. In~this case $\mu^{yx}$ and $\mu^{uu}$ cannot be normalized and the structure equations of the coframe $\{\omega^x, \omega^y, \mu^{yx}, \mu^{uu}\}$ are
\begin{gather*}
{\rm d}\omega^x = \frac{1}{2} \mu^{uu}\wedge \omega^x - \epsilon \mu^{yx} \wedge \omega^y,\quad\
{\rm d}\omega^y = \mu^{yx} \wedge \omega^x + \frac{1}{2}\mu^{uu}\wedge \omega^y,\quad\
{\rm d}\mu^{yx} = 0,\quad\
{\rm d}\mu^{uu} = 0.
\end{gather*}

\subsection{Case H.3}

In this section we assume that we are at a hyperbolic point where $\epsilon=-1$. Also, we are working under the consideration that $U_{Y^3} \equiv - U_{X^3}\neq 0$. Thus, it is possible to normalize $U_{X^3}=1$. At~order 4, the recurrence relation for $U_{Y^3} + U_{X^3}\equiv 0$, yields the equalities
\begin{gather*}
U_{XY^3} \equiv - U_{X^4} - 3U_{X^2Y^2} - 3U_{X^3Y}, \qquad
U_{Y^4} \equiv 3U_{X^4}+6U_{X^2Y^2} + 8U_{X^3Y}.
\end{gather*}

{\samepage
Thus, $U_{X^4}$, $U_{X^3Y}$, and $U_{X^2Y^2}$ are functionally independent partially normalized invariants. Introducing
\begin{gather*}
\begin{bmatrix}
A_1 \\ A_2 \\ A_3
\end{bmatrix} =
\begin{bmatrix}
1 & 2 & 3 \\
1 & 4 & 3 \\
1 & 2 & 1
\end{bmatrix}
\begin{bmatrix}
U_{X^4} \\ U_{X^3Y} \\ U_{X^2Y^2}
\end{bmatrix}\!,
\end{gather*}
we have that
\begin{gather*}
{\rm d}A_k = -\frac{k}{3} A_k \mu^{uu} \mod(\omega^x,\omega^y),
\end{gather*}
for $k=1,2,3$. We~now need to consider the cases
\begin{gather*}
 {\bf H.3.1}\colon\ A_1^2+A_2^2+A_3^2\neq 0,\qquad
 {\bf H.3.2}\colon\ A_1\equiv A_2\equiv A_3\equiv 0.
\end{gather*}
}

Before considering each case, we note that coordinate expressions for the invariants $A_i$ can be found using the method of recursive moving frame. We~obtained
\begin{gather}
A_1 = \frac{2\sqrt{|h|}(2C_2h_x-hC_{2,x}) + u_{xy}(C_2h_x-2hC_{2,x}) + u_{xx}(2hC_{2,y} - C_2h_y)}{\sqrt[3]{2}|h|^{7/6}C_2^{4/3}a_{33}^{1/3}},\nonumber
\\
A_2 = \frac{\sqrt[3]{2}(u_{xx}(hC_{2,y} - 2C_2h_y)+(hC_{2,x}-2C_2h_x)(\sqrt{|h|}-u_{xy}))}{u_{xx}|h|^{11/6}(C_2a_{33})^{2/3}},\nonumber
\\
A_3 = \frac{1}{8h^3 u_{xx}^2 a_{33}}\big(\sqrt{|h|}u_{xx}(3C_2h_y - 2hC_{2,y})+\sqrt{|h|}u_{xy}(2hC_{2,x}-3C_2h_x)\nonumber
\\ \hphantom{A_3 =}
{}+h(2hC_{2,x}-3C_2h_x)+4h^2h_{xx}u_{xx}-h_y^2u_{xx}^3+2h_xh_yu_{xx}^2u_{xy}\nonumber
\\ \hphantom{A_3 =}
{}-h_x^2u_{xx}u_{xy}^2-6hh_x^2u_{xx}\big).
\label{eq:A invariants}
\end{gather}

\subsubsection{Case H.3.1}

In this case there is $A_k$, with $k\in \{1,2,3\}$, such that $A_k\neq 0$. For the sake of the exposition, assume $A_3 \neq 0$. The other possibilities are dealt in a similar fashion. When $A_3\neq 0$, one can normalize $A_3=1$. Then
\begin{gather*}
{\rm d}A_1 = \frac{1}{3}\big[(3-A_1)U_{X^5} +2(3-A_1)U_{X^4Y}+(9-A_1) U_{X^3Y^2}-6-2A_1^2+A_1A_2+12A_2\big]\omega^x
\\ \hphantom{{\rm d}A_1 =}
{}+\frac{1}{3}\big[(A_1\!-\!9)U_{X^5} \!+\!2(A_1\!-\!12)U_{X^4Y}'! +\! (A_1\!-\!21)U_{X^3Y^2}\!-6 \!+\!2 A_1^2\!+\! A_1 A_2 \!- \!12 A_2\big]\omega^y,
\\
{\rm d}A_2 = \frac{1}{6}\big[2(3-2A_2)U_{X^5}+8(3-A_2)U_{X^4Y}+2(9-2A_2)U_{X^3Y^2}+42-2A_1A_2 + A_2^2\big]\omega^x
\\ \hphantom{{\rm d}A_2 =}
{}+\frac{1}{6}\big[2(2A_2\!-\!9)U_{X^5} \!+\!8(A_2\!-\!6)U_{X^4Y} \!+\! 2(2A_2\!-\!15)U_{X^3Y^2}\!-\!42\!+\!2A_1A_2 \!+\! A_2^2\big]\omega^y,
\end{gather*}
and we have the structure equations
\begin{gather*}
{\rm d}\omega^x = \frac{1}{12}(8I-2A_1+A_2)\omega^x\wedge \omega^y, \qquad
{\rm d}\omega^y = \frac{1}{12}(8I-2A_1-A_2)\omega^x\wedge \omega^y,
\end{gather*}
where $I= U_{X^5}+2U_{X^4Y} + U_{X^3Y^2}$. Since
\begin{gather*}
\D_y A_2 + \D_x A_2 = \frac{1}{3}A_2^2-2I,
\end{gather*}
it follows that $I$ can be expressed in terms of $A_2$ and its invariant derivatives. From the syzygy
\begin{gather*}
A_1(6I-A_2^2) =\frac{A_2^3}{2} - 6I A_2^2 + +3A_2(4\D_xI + 2I + 4 \D_yI-5\D_xA_2)
\\ \hphantom{A_1(6I-A_2^2) =}
{}+ 6\big(4I^2-3\D_xI - 9 \D_yI+3\D_x^2A_2 + 3\D_y\D_xA_2\big),
\end{gather*}
it follows that $A_1$ can generically be expressed in terms of $A_2$ and its invariant derivatives.

\begin{Theorem}
The algebra of differential invariants is generically generated by the single inva\-riant $A_2$.
\end{Theorem}

\begin{Remark}
Solving the normalization equation $A_3=1$ we obtain
\begin{gather*}
a_{33} = \frac{1}{8h^3 u_{xx}^2}\big(\sqrt{|h|}u_{xx}(3C_2h_y-2hC_{2,y})+\sqrt{|h|}u_{xy}(2hC_{2,x}-3C_2h_x)
\\ \hphantom{a_{33} =}
{}+h(2hC_{2,x}-3C_2h_x)+4h^2h_{xx}u_{xx}-h_y^2u_{xx}^3+2h_xh_yu_{xx}^2u_{xy}-h_x^2u_{xx}u_{xy}^2-6hh_x^2u_{xx}\big).
\end{gather*}
Substituting this group parameter normalization into the formula for $A_2$ in~\eqref{eq:A invariants} yields the coordinate expression for the generating invariant $A_2$.
\end{Remark}

\subsubsection{Case H.3.2}

When $A_1\equiv A_2 \equiv A_3 \equiv 0$, there is no further group parameter normalizations possible. Then, the structure equations of the coframe $\{\omega^x,\omega^y,\mu^{uu}\}$ are
\begin{gather}\label{eq:H.3.2 structure}
{\rm d}\omega^x = \frac{1}{2} \mu^{uu}\wedge \omega^x + \frac{1}{6} \mu^{uu}\wedge \omega^y,\qquad
{\rm d}\omega^y = \frac{1}{6} \mu^{uu}\wedge \omega^x + \frac{1}{2} \mu^{uu}\wedge \omega^y,\qquad {\rm d}\mu^{uu}=0.
\end{gather}

\section{Parabolic points}\label{sec: parabolic}

At a parabolic point, $H = U_{X^2}U_{Y^2} - U_{XY}^2 \equiv 0$. Therefore,
\begin{gather}\label{eq:H=0}
U_{XY}^2 \equiv U_{X^2}U_{Y^2}
\end{gather}
and there are now two cases to consider. Namely,
\begin{gather*}
{\bf P.1}\colon\ U_{X^2}\neq 0,\qquad
{\bf P.2}\colon\ U_{X^2}\equiv 0.
\end{gather*}

\subsection{Case P.1}

When $U_{X^2}\neq 0$, we can solve for $U_{Y^2}$ in~\eqref{eq:H=0} to obtain
\begin{gather}\label{eq:UY2}
U_{Y^2} \equiv \frac{U_{XY}^2}{U_{X^2}}.
\end{gather}
Therefore, $U_{X^2}$ and $U_{XY}$ are functionally independent partially normalized invariants. From the recurrence relations~\eqref{eq: order 2 recurrence relations}, we conclude that it is possible to set
\begin{gather}\label{eq:P.1 order 2 normalizations}
U_{XX}=1,\qquad U_{XY}=0,
\end{gather}
and~\eqref{eq:UY2} implies that $U_{YY} \equiv 0$. Taking into account the normalizations~\eqref{eq:P.1 order 2 normalizations} and the equality $U_{YY}\equiv 0$, the recurrence relation for $U_{YY}$ in~\eqref{eq: order 2 recurrence relations} implies that
\begin{gather}\label{eq:P.1 order 3 constraints}
U_{XY^2} \equiv U_{Y^3} \equiv 0,
\end{gather}
which in turn yields the recurrence relations
\begin{align*}
0 &\equiv {\rm d}U_{XY^2} = \big({-}2U_{X^2Y}^2 + U_{X^2Y^2}\big)\omega^x + U_{XY^3}\omega^y,\\
0 &\equiv {\rm d}U_{Y^3} = U_{XY^3}\omega^x + U_{Y^4}\omega^y.
\end{align*}
Therefore,
\begin{gather}\label{eq:order 4 constraints}
U_{Y^4}\equiv U_{XY^3}\equiv 0,\qquad U_{X^2Y^2}\equiv 2U_{X^2Y}^2.
\end{gather}
Considering the recurrence relations for the third order partially normalized invariants $U_{X^3}$ and $U_{X^2Y}$, and taking into account the above constraints on the invariants, we find that
\begin{gather*}
{\rm d}U_{X^3} = - 3\mu^{xu} - \frac{1}{2}U_{X^3}\mu^{uu}-3U_{X^2Y}\mu^{yx} \mod(\omega^x,\omega^y),
\\
{\rm d}U_{X^2Y} = -U_{X^2Y}\mu^{yy} \mod(\omega^x,\omega^y).
\end{gather*}
From the first equation we conclude that it is possible to normalize $U_{X^3}=0$. As for the second recurrence relation, we have the following cases to consider
\begin{gather*}
{\bf P.1.1}\colon\ U_{X^2Y}\neq 0,\qquad
{\bf P.1.2}\colon\ U_{X^2Y}\equiv 0.
\end{gather*}
Introducing the ratio $R= \cfrac{u_{xy}}{u_{xx}}$, the coordinate expression for $U_{X^2Y}$ is
\begin{gather*}
U_{X^2Y} = \frac{R_x}{Y_y-RY_x},
\end{gather*}
where we recall that $Y_x$ and $Y_y$ are introduced in~\eqref{eq:det}.

\subsubsection{Case P.1.1}

When $U_{X^2Y}\neq 0$, we can normalize $U_{X^2Y}=1$. Then the recurrence relations for the non-constant fourth order partially normalized lifted invariants, i.e., $U_{X^4}$ and $U_{X^3Y}$, are
\begin{gather}
{\rm d}U_{X^4} = - 6\mu^{yu} - U_{X^4}\mu^{uu}-4U_{X^3Y}\mu^{yx}\mod(\omega^x,\omega^y),\nonumber
\\
{\rm d}U_{X^3Y} = -\frac{1}{2}U_{X^3Y}\mu^{uu}
 \mod(\omega^x,\omega^y).
\label{eq:P.1.1 recurrence relations}
\end{gather}
From the first equation, we see that it is possible to normalize $U_{X^4}=0$. Next, the exterior derivative of the constraints~\eqref{eq:order 4 constraints} yields
\begin{gather*}
0 = {\rm d}U_{X^2Y^2} = \big(U_{X^3Y^2}-6U_{X^3Y}\big)\omega^x+(U_{X^2Y^3}-6)\omega^y,\\
0 \equiv {\rm d}U_{XY^3} = \big(U_{X^2Y^3}-6\big)\omega^x + U_{XY^4}\omega^y,\\
0 \equiv {\rm d}U_{Y^4} = U_{XY^4}\omega^x + U_{Y^5}\omega^y,
\end{gather*}
from which we obtain the following constraints among the order 5 partially normalized invariants
\begin{gather}\label{eq:P.1.1 constraints}
U_{Y^5}\equiv U_{XY^4} \equiv 0,\qquad U_{X^2Y^3}\equiv 6,\qquad U_{X^3Y^2} = 6U_{X^3Y}.
\end{gather}
In light of the second equation in~\eqref{eq:P.1.1 recurrence relations}, we now have to consider the following cases
\begin{gather*}
{\bf P.1.1.1}\colon\ U_{X^3Y}\neq 0,\qquad
{\bf P.1.1.2}\colon\ U_{X^3Y}\equiv 0,
\end{gather*}
where
\begin{gather*}
U_{X^3Y} = \frac{R_{xx}}{R_x \sqrt{|a_{33}u_{xx}}|}.
\end{gather*}

\subsubsection{Case P.1.1.1}

When $U_{X^3Y}\neq 0$, we set $U_{X^3Y}=1$. Then the recurrence relation for $U_{X^4Y}$ is
\begin{gather*}
{\rm d}U_{X^4Y} = -2\mu^{yx} \mod (\omega^x,\omega^y),
\end{gather*}
and so we can normalize $U_{X^4Y}=0$. At this stage, the recurrence relation for the only remaining fifth order normalized invariant is
\begin{gather}\label{eq:P.1.1.1 recurrence relation}
{\rm d}U_{X^5} = \frac{1}{3}\big(3U_{X^6}+10U_{X^5Y}-36U_{X^5}\big)\omega^x+\frac{1}{9}\big(U_{X^5Y}+80-63U_{X^5}\big)\omega^y.
\end{gather}
The exterior derivative of the constraints~\eqref{eq:P.1.1 constraints} yields
\begin{gather*}
0 \equiv {\rm d}U_{X^3Y^2} = \big(U_{X^4Y^2}-6\big)\omega^x+\big(U_{X^3Y^3}-36\big)\omega^y,\\
0 \equiv {\rm d}U_{X^2Y^3} = \big(U_{X^3Y^3}-36\big)\omega^x+\big(U_{X^2Y^4}-24\big)\omega^y,\\
0 \equiv {\rm d}U_{XY^4} = \big(U_{X^2Y^4}-24\big)\omega^x+U_{XY^5}\omega^y,\\
0 \equiv {\rm d}U_{Y^5} = U_{XY^5}\omega^x + U_{Y^6}\omega^y.
\end{gather*}
Thus
\begin{gather*}
U_{Y^6}=0,\qquad U_{XY^5}=0,\qquad U_{X^2Y^4}=24,\qquad U_{X^3Y^3}=36,\qquad
U_{X^4Y^2}=6,
\end{gather*}
and it follows that $U_{X^6}$ and $U_{X^5Y}$ are the only functionally independent invariants of order 6.

From~\eqref{eq:P.1.1.1 recurrence relation} we conclude that $U_{X^5Y}$ and $U_{X^6}$ can be expressed in terms of $U_{X^5}$ and its invariant derivatives. It follows from Theorem \ref{thm:generating set} that $U_{X^5}$ generates the algebra of differential invariants. Introducing the ratios
\begin{gather*}
S = \frac{3u_{xx} u_{xxxx} - 5 u_{xxx}^2}{3u_{xx}^2}\qquad\text{and}\qquad L=\frac{u_{xxx}}{u_{xx}},
\end{gather*}
we have that
\begin{gather*}
U_{X^5} = \frac{R_x}{36R_{xx}^4}\big(30 L R_x R_{xx} R_{xxx} -24 L S R_x^2 R_{xx} - 5L^2R_x R_{xx}^2 - 60 S R_x R_{xx}^2 - 40 L R_{xx}^3
\\ \hphantom{U_{X^5} =}
{}+ 120 R_{xx}^2 R_{xxx} - 45R_xR_{xxx}^2 + 36 R_x^2 R_{xx} S_x\big).
\end{gather*}
Finally, the structure equations of the invariant coframe $\{\omega^x,\omega^y\}$ are
\begin{gather*}
{\rm d}\omega^x = \omega^y \wedge \omega^x,\qquad {\rm d}\omega^y = \frac{1}{3}\omega^y\wedge\omega^x.
\end{gather*}

\subsubsection{Case P.1.1.2}

When $U_{X^3Y}\equiv 0$, $0\equiv {\rm d}U_{X^3Y} = U_{X^4Y}\omega^x + U_{X^3Y^2}\omega^y$, which, when combined with~\eqref{eq:P.1.1 constraints}, implies that
\begin{gather}\label{eq:P.1.1.2 constraints}
U_{Y^5}\equiv U_{XY^4}\equiv U_{X^3Y^2} \equiv U_{X^4Y}\equiv 0,\qquad U_{X^2Y^3}\equiv 6.
\end{gather}
Thus, the recurrence relation for the only non-constant order 5 partially normalized invariant, namely
\begin{gather*}
U_{X^5} = \frac{T}{3\sqrt{|a_{33}u_{xx}|}^3},
\end{gather*}
where $T=2LS-3S_x$, is
\begin{gather}\label{eq:dX5}
{\rm d}U_{X^5} =-\frac{3}{2}U_{X^5}\mu^{uu} \mod(\omega^x,\omega^y).
\end{gather}
Next, the recurrence relations for the constant invariants~\eqref{eq:P.1.1.2 constraints} are
\begin{gather*}
0 \equiv {\rm d}U_{X^4Y} = \big(U_{X^4Y}-4U_{X^5}\big)\omega^x+U_{X^4Y^2}\omega^y,\\
0 \equiv {\rm d}U_{X^3Y^2} = U_{X^4Y^2}\omega^x+U_{X^3Y^3}\omega^y,\\
0 \equiv {\rm d}U_{X^2Y^3} = U_{X^3Y^3}\omega^x+\big(U_{X^2Y^4}-24\big)\omega^y,\\
0 \equiv {\rm d}U_{XY^4} = \big(U_{X^2Y^4}-24\big)\omega^x+U_{XY^5}\omega^y,\\
0 \equiv {\rm d}U_{Y^5} = U_{XY^5}\omega^x + U_{Y^6}\omega^y.
\end{gather*}
These equations imply that
\begin{gather}\label{eq:P.1.1.2 order 6 constraints}
U_{Y^6}\equiv U_{XY^5}\equiv U_{X^3Y^3}\equiv U_{X^4Y^2}\equiv 0,\qquad U_{X^2Y^4}\equiv 24,\qquad U_{X^5Y}\equiv 4U_{X^5}.
\end{gather}
In light of~\eqref{eq:dX5}, we have the following cases to consider
\begin{gather*}
{\bf P.1.1.2.1}\colon\ U_{X^5}\neq 0,\qquad
{\bf P.1.1.2.2}\colon\ U_{X^5}\equiv 0.
\end{gather*}

\subsubsection{Case P.1.1.2.1}

In this case we normalize $U_{X^5}=1$. Then the recurrence relation for $U_{X^6}$ is
\begin{gather*}
{\rm d}U_{X^6}=-3\mu^{yx} \mod(\omega^x,\omega^y),
\end{gather*}
and it is therefore also possible to set $U_{X^6}=0$. At this stage all invariants of order 6 or less are constant and the only non-constant invariant of order 7 is
\begin{gather*}
U_{X^7} = -\frac{1}{6}-\frac{3^{2/3}L^2}{2T^{2/3}} - \frac{3^{5/3}S}{2T^{2/3}}-\frac{7T_x^2}{2\cdot 3^{1/3}T^{8/3}}+\frac{3^{2/3}T_{xx}}{T^{5/3}}.
\end{gather*}
Similarly, the only non-phantom invariant of order 8 is $U_{X^8}$. From the recurrence relation
\begin{gather*}
{\rm d}U_{X^7} = \bigg(U_{X^8}-\frac{35}{2}\bigg)\omega^x
\end{gather*}
it follows that $U_{X^8} = \D_xU_{X^7}+\frac{35}{2}$, and from Theorem \ref{thm:generating set}, $U_{X^7}$ generates the algebra of differential invariants. Finally, the structure equations are
\begin{gather*}
{\rm d}\omega^x = 0,\qquad {\rm d}\omega^y=\frac{5}{3}\omega^y\wedge\omega^x.
\end{gather*}

\subsubsection{Case P.1.1.2.2}

When $U_{X^5} \equiv 0$, we have that
\begin{gather*}
0 \equiv {\rm d}U_{X^5} = U_{X^6}\omega^x,
\end{gather*}
which when combined with~\eqref{eq:P.1.1.2 order 6 constraints}, implies that all sixth order invariants are constant. Similarly, all higher order invariants are constant and there are no further possible normalizations. Finally, the structure equations for the coframe $\{\omega^x,\omega^y,\mu^{yx},\mu^{uu}\}$ are
\begin{gather*}
{\rm d}\omega^x=\frac{1}{2}\omega^x\wedge \omega^y,\qquad
{\rm d}\omega^y = \mu^{yx}\wedge \omega^x,
\\
{\rm d}\mu^{yx} = \frac{1}{2}\mu^{yx}\wedge \omega^y + \frac{1}{2}\mu^{uu}\wedge \mu^{yx},\qquad
{\rm d}\mu^{uu} = \mu^{yx}\wedge \omega^x.
\end{gather*}

\subsubsection{Case P.1.2}

When $U_{X^2Y}\equiv 0$, we have $U_{X^2Y}\equiv U_{XY^2}\equiv U_{Y^3}\equiv0$, in light of~\eqref{eq:P.1 order 3 constraints}. We~also recall that $U_{X^3}$ is normalized to zero. From the recurrence relations
\begin{gather*}
0 \equiv {\rm d}U_{X^2Y} = U_{X^3Y}\omega^x + U_{X^2Y^2}\omega^y, \\
0 \equiv {\rm d}U_{XY^2} = U_{X^2Y^2}\omega^x + U_{XY^3}\omega^y, \\
0 \equiv {\rm d}U_{Y^3} = U_{XY^3}\omega^x + U_{Y^4}\omega^y,
\end{gather*}
we conclude that
\begin{gather}\label{eq:P.1.2 order 4 constraints}
U_{XY^3} \equiv U_{X^2Y^2}\equiv U_{X^3Y}\equiv U_{Y^4} \equiv 0.
\end{gather}
Thus,
\begin{gather*}
U_{X^4} = \frac{S}{a_{33}u_{xx}}
\end{gather*}
is the lowest order non-zero invariant and the recurrence relation
\begin{gather*}
{\rm d}U_{X^4}= - U_{X^4}\mu^{uu} \mod (\omega^x,\omega^y),
\end{gather*}
leads us to consider the following cases
\begin{gather*}
{\bf P.1.2.1}\colon\ U_{X^4}\neq 0,\qquad
{\bf P.1.2.2}\colon\ U_{X^4}\equiv 0.
\end{gather*}

\subsubsection{Case P.1.2.1}

In this case we can normalize $U_{X^4}=1$. Since the recurrence relations for the vanishing invariants~\eqref{eq:P.1.2 order 4 constraints} are of the form $0\equiv {\rm d}U_J = U_{J,i}\omega^i$, all fifth order partially normalized invariants are zero except for $U_{X^5}$. Similarly, all sixth order partially normalized invariants are zero except for~$U_{X^6}$. Since
\begin{gather*}
{\rm d}U_{X^5}=-\frac{1}{2}\big(10+3U_{X^5}^2-2U_{X^6}\big) \omega^x,
\end{gather*}
the function $U_{X^5}$ is a \emph{genuine} differential invariant not depending on the remaining group parameters. By a similar argument, we see that for $k\geq 5$, $U_{X^k}$ are genuine differential invariants, while $U_{X^kY^\ell}\equiv 0$ for $\ell>0$ and $k+\ell \geq4$. It follows that
\begin{gather*}
U_{X^5}=\frac{3T}{S^{3/2}}
\end{gather*}
generates the algebra of differential invariants.

When $U_{X^5}=c$ is constant, it follows that $U_{X^k}$, $k\geq 5$, are all constant and the symmetry group of these surfaces has structure equations
\begin{gather*}
{\rm d}\omega^x=0,\qquad
{\rm d}\omega^y=\mu^{yx}\wedge \omega^x +\mu^{yy}\wedge \omega^y,
\\
{\rm d}\mu^{yu} = c \omega^x\wedge \mu^{yu}+\frac{1}{3}\omega^x\wedge \mu^{yx}+\mu^{yu}\wedge \mu^{yy},\\
{\rm d}\mu^{yx} = \frac{c}{2} \omega^x\wedge \mu^{yx}+\mu^{yu}\wedge \omega^x+\mu^{yx}\wedge \mu^{yy},\qquad
 {\rm d}\mu^{yy}= 0.
\end{gather*}

\subsubsection{Case P.1.2.2}

If $U_{X^4}\equiv 0$, then in light of~\eqref{eq:P.1.2 order 4 constraints} all fourth order partially normalized invariants are zero and there are no non-trivial invariants. These surfaces have a symmetry group with structure equations
\begin{gather*}
{\rm d}\omega^x=\frac{1}{2}\mu^{uu}\wedge\omega^x,\qquad
{\rm d}\omega^y = \mu^{yx} \wedge \omega^x+\mu^{yy}\wedge \omega^y,\qquad {\rm d}\mu^{uu}=0,\qquad {\rm d}\mu^{yy}=0,\\
{\rm d}\mu^{yu}=\mu^{yu}\wedge \mu^{yy}+\mu^{uu}\wedge \mu^{yu},\qquad
{\rm d}\mu^{yx}=\frac{1}{2}\mu^{uu}\wedge \mu^{yx}+\mu^{yx}\wedge \mu^{yy}+\mu^{yu}\wedge \omega^x.
\end{gather*}

\subsection{Case P.2}

If $U_{X^2} \equiv 0$, then equation~\eqref{eq:H=0} implies that $U_{XY} \equiv 0$. Since
\begin{gather*}
0={\rm d}U_{XY}=-U_{YY}\mu^{yx} \mod(\omega^x\wedge \omega^y),
\end{gather*}
it follows that $U_{YY}\equiv 0$. Such surfaces have a 9-dimensional symmetry group with structure equations
\begin{gather*}
{\rm d}\omega^x=\mu^{xx}\wedge \omega^x+\mu^{xy}\wedge \omega^y,\qquad
{\rm d}\omega^y = \mu^{yx}\wedge \omega^x + \mu^{yy}\wedge \omega^y,\qquad {\rm d}\mu^{uu}=0,\\
{\rm d}\mu^{xx} = \mu^{yx}\wedge \mu^{xy},\qquad
{\rm d}\mu^{xy} = \mu^{xy}\wedge \mu^{xx}+\mu^{yy}\wedge \mu^{xy},\\
{\rm d}\mu^{xu}=\mu^{xu}\wedge \mu^{xx} + \mu^{yu}\wedge \mu^{xy}+\mu^{uu}\wedge \mu^{xu},\qquad
{\rm d}\mu^{yx}=\mu^{xx}\wedge \mu^{yx}+\mu^{yx}\wedge \mu^{yy},\\
{\rm d}\mu^{yy}=\mu^{xy}\wedge \mu^{yx},\qquad
{\rm d}\mu^{yu} = \mu^{xu}\wedge\mu^{yx}+\mu^{yu}\wedge \mu^{yy}+\mu^{uu}\wedge\mu^{yu}.
\end{gather*}

\section{Homogeneous surfaces}\label{sec:homogeneous surfaces}

As mentioned in the introduction, differential geometers have been especially interested in the study of homogeneous surfaces that arise from the equivalence problem~\cite{ADV-1997,DKR-1996,EE-1999}. These surfaces are characterized by the property that all relative and differential invariants are constant. Therefore, homogeneous surfaces are described as solutions to certain systems of partial differential equations. We~now consider several examples, with the understanding that it is not our intention to recover the full classifications found in~\cite{ADV-1997,DKR-1996,EE-1999}.

\begin{Example}
As our first example, let us consider the branch EH.2.1. Surfaces belonging to this branch satisfy the partial differential equations $U_{X^3}\equiv U_{Y^3} \equiv 0$ and the non-degeneracy condition $U_{X^2Y^2}\neq 0$. In~jet coordinates, these conditions translate to the formulas
\begin{gather}\label{eq:H.2.1 homogeneous surface PDE}
u_ {xyy}=\frac{u_{xxx} \big(u_{xx} u_{yy}-4 u_ {xy}^2\big)+6 u_{xx} u_{xy}u_{xxy}}{3 u_ {xx}^2},\qquad
u_ {yyy}=\frac{u_{yy} (3 u_{xx} u_{xxy}-2u_{xxx} u_{xy})}{u_ {xx}^2},
\end{gather}
and
\begin{gather*}
\big(4 u_{xxx}^2\!+3 u_{xx} u_{xxxx}\big) u_{xy}^2-18 u_{xx} u_{xy} u_{xxx}u_{xxy} \!+9 u_{xx}^2 u_{xxy}^2\!+u_{xx}u_{yy}\big(5 u_{xxx}^2\!-3 u_{xx}u_{xxxx}\big) \neq0.
\end{gather*}
Our results say that all surfaces satisfying this system are equivalent, and each is a homogeneous space with symmetry group of dimension~3. A normal form for this branch can therefore be taken as any solution to the above system. Completing~\eqref{eq:H.2.1 homogeneous surface PDE} to an involutive system~\cite{S-2010}, one obtains a \emph{maximally overdetermined} fifth order system, which can be solved using the Frobenius theorem. We~find that the non-degenerate quadrics
\begin{gather*}
\frac{x^2}{a^2}+\frac{y^2}{b^2}\pm \frac{u^2}{c^2}=1,\qquad
\frac{x^2}{a^2}+\frac{y^2}{b^2}-\frac{u^2}{c^2}=-1,
\end{gather*}
with $c\neq 0$, satisfy the constraints of this branch. These surfaces correspond to case (1) in \cite[Theorem 1.1]{ADV-1997}.
\end{Example}

\begin{Example}
In Case EH.2.2 we also obtain homogeneous surfaces since all the invariants are constant. In~this case, the surface must be a solution to the system of differential equations
\begin{gather*}
u_ {xyy}=\frac{u_{xxx} \big(u_{xx} u_{yy}-4 u_ {xy}^2\big)+6 u_{xx} u_{xy}u_{xxy}}{3 u_ {xx}^2},\qquad
u_ {yyy}=\frac{u_{yy} (3 u_{xx} u_{xxy}-2u_{xxx} u_{xy})}{u_ {xx}^2},\\
\big(4 u_{xxx}^2\!+3 u_{xx} u_{xxxx}\big) u_{xy}^2\!-18 u_{xx}u_{xy} u_{xxx}u_{xxy} \!+9 u_{xx}^2 u_{xxy}^2\!+u_{xx}u_{yy}\big(5 u_{xxx}^2\!-3 u_{xx}u_{xxxx}\big) = 0.
\end{gather*}
One can verify that the remain two non-degenerate quadrics $u=\frac{y^2}{b^2}\pm \frac{x^2}{a^2}$ are solutions. Setting $a=b=1$, we recover case (8) of \cite[Theorem 1]{DKR-1996} with $\alpha=2$.
\end{Example}

\begin{Example}
In Case P.1.1.2.2, the homogeneous surface must satisfy the system of differential equations
\begin{gather*}
u_{xx} u_{yy} - u_{xy}^2 =0,\qquad
45u_{xx} u_{xxx} u_{xxxx} - 9 u_{xx}^2 u_{xxxxx} -40 u_{xxx}^3=0,\\
u_{xx}^2 u_{xxxy}-u_{xx}u_{xy}u_{xxxx}-2u_{xx}u_{xxy} u_{xxx}+2u_{xy}u_{xxx}^2=0,
\end{gather*}
and the non-degeneracy conditions
\begin{gather*}
u_{xx}\neq 0,\qquad u_{xx}u_{xxy}-u_{xy}u_{xxx}\neq 0.
\end{gather*}
A solution is given by $u=x^2 y^{-1}$, corresponding to case (1) of \cite[Theorem 1]{DKR-1996} with $\alpha=2$ and~$\beta=-1$.
\end{Example}

\begin{Example}
A homogeneous surface in branch P.1.2.2 must satisfy the system of partial differential equations
\begin{gather*}
u_{xx}u_{yy}-u_{xy}^2=0,\qquad u_{xxy}u_{xx}-u_{xy}u_{xxx}=0,\qquad 3u_{xx}u_{xxxx}-5u_{xxx}^2=0.
\end{gather*}
A solution is given by $u=x^2$, corresponding to case (1) of \cite[Theorem 1]{DKR-1996}.
\end{Example}

\begin{Example}
A homogeneous surface in branch P.2 will be a solution to the system of dif\-fe\-ren\-tial equations
\begin{gather*}
u_{xx}=u_{xy}=u_{yy}=0.
\end{gather*}
The general solution being a plane $u=ax+by+c$.
\end{Example}

The above examples show that attempting to recover the homogeneous surfaces from the systems of partial differential equations one obtains by setting the relative or differential invariants to constant values can be extremely challenging as these equations are highly nonlinear and of high order. Luckily, it is possible to avoid these difficulties by integrating the moving frame equations instead~\cite{G-1963}. To~see how this works, let $\widehat{\rho}\n=\big(\rho\n,z\n\big)$ be a partial right moving frame. As is customary, we also refer to $\rho\n=\rho\n \in G$ as a partial right moving frame. Then let $\overline{\rho}\n = \big(\rho\n\big)^{-1}$ denote the partial left moving frame. Taking the exterior derivative of the identity $\rho\n\overline{\rho}\n = \mathds{1}$, we find that
\begin{gather}\label{eq:reconstruction}
{\rm d}\overline{\rho}\n = -\overline{\rho}\n \bmu^*,
\end{gather}
where $\bmu^*=\big(\rho\n\big)^*\bmu$ denotes the right moving frame pull-back of the Maurer--Cartan forms. To~proceed further, let
\begin{gather*}
\overline{\rho}\n = \begin{bmatrix}
E & z \\
0 & 1
\end{bmatrix}\!,
\end{gather*}
where $E=(\ve_1 \ve_2 \ve_3)\in GL(3)$ is a frame on the homogeneous surface $S\subset \mathbb{R}^3$ and $z\in \mathbb{R}^3$ is a~point in $S$. It therefore follows that if one can integrate the moving frame equation~\eqref{eq:reconstruction} for~$\overline{\rho}\n$, a parametrization of the homogeneous surface will be given by the vector $z\in \mathbb{R}^3$. We~now show how this works with a concrete example.

\begin{Example}
In this example we will deduce the homogeneous surface that originates from Case H.3.2. Recall the structure equations obtained in~\eqref{eq:H.3.2 structure}. From the third equation, it follows that, locally,
\begin{gather*}
\mu^{uu} ={\rm d}a.
\end{gather*}
Next, introduce
\begin{gather}\label{eq:overline omega}
\overline{\omega}^x = {\rm e}^{-2a/3}(\omega^x+\omega^y),\qquad
\overline{\omega}^y = {\rm e}^{-a/3}(\omega^x-\omega^y).
\end{gather}
Using~\eqref{eq:H.3.2 structure} we find that
\begin{gather*}
{\rm d}\overline{\omega}^x = {\rm d}\overline{\omega}^y = 0.
\end{gather*}
Therefore, locally,
\begin{gather}\label{eq:oxy}
\overline{\omega}^x = 2 {\rm d}\ox,\qquad
\overline{\omega}=2 {\rm d}\oy,
\end{gather}
for certain functions $\ox$ and $\oy$. Substituting~\eqref{eq:oxy} in~\eqref{eq:overline omega} and solving for $\omega^x$ and $\omega^y$ we obtain
\begin{gather*}
\omega^x = {\rm e}^{2a/3}{\rm d}\ox + {\rm e}^{a/3}{\rm d}\oy,\qquad
\omega^y = {\rm e}^{2a/3}{\rm d}\ox - {\rm e}^{a/3}{\rm d}\oy.
\end{gather*}
Since
\begin{gather}
\bmu^*=
\begin{bmatrix}
\frac{1}{2}(\omega^x-\omega^y)+\frac{1}{2}\mu^{uu} & -\frac{1}{2}(\omega^x-\omega^y)+\frac{1}{6}\mu^{uu} & 0 & -\omega^x \\[1ex]
\frac{1}{2}(\omega^x-\omega^y) +\frac{1}{6}\mu^{uu} & -\frac{1}{2}(\omega^x-\omega^y)+\frac{1}{2}\mu^{uu} & 0 & -\omega^y \\
-\omega^x & \omega^y &\mu^{uu} & 0 \\
0 & 0 & 0 & 0
\end{bmatrix} \nonumber
\\ \hphantom{\bmu^*}
=\begin{bmatrix}
{\rm e}^{a/3}{\rm d}\oy + \frac{1}{2}{\rm d}a & -a^{a/3}{\rm d}\oy + \frac{1}{6}{\rm d}a & 0 & -{\rm e}^{2a/3}{\rm d}\ox - {\rm e}^{a/3}{\rm d}\oy \\[1ex]
{\rm e}^{a/3}{\rm d}\oy + \frac{1}{6}{\rm d}a & -{\rm e}^{a/3}{\rm d}\oy + \frac{1}{2}{\rm d}a & 0 & -{\rm e}^{2a/3}{\rm d}\ox + {\rm e}^{a/3}{\rm d}\oy \\[1ex]
-{\rm e}^{2a/3}{\rm d}\ox - {\rm e}^{a/3}{\rm d}\oy & {\rm e}^{2a/3}{\rm d}\ox - {\rm e}^{a/3}{\rm d}\oy & {\rm d}a & 0 \\
0 & 0 & 0 &0
\end{bmatrix}\!,
\label{bmu}
\end{gather}
equation~\eqref{eq:reconstruction} yields
\begin{gather*}
{\rm d}z = \big({\rm e}^{2a/3}{\rm d}\ox + {\rm e}^{a/3}{\rm d}\oy\big)\ve_1
+ \big({\rm e}^{2a/3}{\rm d}\ox - {\rm e}^{a/3}{\rm d}\oy\big)\ve_2,
\\
{\rm d}\ve_1 = -\bigg({\rm e}^{a/3} {\rm d}\oy + \frac{{\rm d}a}{2}\bigg)\ve_1 - \bigg({\rm e}^{a/3}{\rm d}\oy
+ \frac{{\rm d}a}{2}\bigg)\ve_2 + \big({\rm e}^{2a/3}{\rm d}\ox + {\rm e}^{a/3}{\rm d}\oy\big)\ve_3,
\\
{\rm d}\ve_2 = -\bigg({-}{\rm e}^{a/3} {\rm d}\oy + \frac{{\rm d}a}{6}\bigg)\ve_1 - \bigg({-}{\rm e}^{a/3} {\rm d}\oy + \frac{{\rm d}a}{2}\bigg)\ve_2 - \big({\rm e}^{2a/3}{\rm d}\ox - {\rm e}^{a/3}{\rm d}\oy\big)\ve_3, \\
{\rm d}\ve_3 = -{\rm d}a \ve_3,
\end{gather*}
from which we conclude that
\begin{gather*}
z_{\ox} = {\rm e}^{2a/3}(\ve_1 + \ve_2), \qquad
z_{\oy} = {\rm e}^{a/3}(\ve_1 - \ve_2),
\\
\ve_{1,\ox} = {\rm e}^{2a/3}\ve_3,\qquad
\ve_{1,\oy} = {\rm e}^{a/3}(-\ve_1-\ve_2+\ve_3),\qquad
\ve_{1,a} = -\frac{1}{2}\ve_1 - \frac{1}{6}\ve_2,
\\
\ve_{2,\ox} = -{\rm e}^{2a/3}\ve_3,\qquad
\ve_{2,\oy} = {\rm e}^{a/3}(\ve_1+\ve_2+\ve_3),\qquad
\ve_{2,a} = -\frac{1}{6}\ve_1 - \frac{1}{2}\ve_2,
\\
\ve_{3,\ox} =\ve_{3,\oy}=0,\qquad \ve_{3,a}=-\ve_3.
\end{gather*}
and
\begin{gather*}
z_{
\ox\ox} = 0,\qquad z_{\ox\oy} = 2e^a\ve_3,\qquad z_{\oy\oy} = -2{\rm e}^{2a/3}(\ve_1+\ve_2),\qquad z_{\oy\oy\oy}= -4e^a\ve_3.
\end{gather*}
Integrating the latter system of equations, we obtain
\begin{gather*}
z = \big({\rm e}^{2a/3}\ox + {\rm e}^{a/3}\oy + {\rm e}^{2a/3}\oy^2\big)\ve_1 + \big({\rm e}^{2a/3}\ox - {\rm e}^{a/3}\oy + {\rm e}^{2a/3}\oy^2\big)\ve_2 - 2e^a \bigg(\ox\oy + \frac{1}{3}\oy^3\bigg)\ve_3.
\end{gather*}
Introducing the variables
\begin{gather*}
x = {\rm e}^{2a/3}\ox + {\rm e}^{a/3}\oy + {\rm e}^{2a/3}\oy^2,\quad\ \
y = {\rm e}^{2a/3}\ox-{\rm e}^{a/3}\oy + {\rm e}^{2a/3}\oy^2,\quad\ \
u= -2e^a\bigg(\ox\oy+\frac{1}{3}\oy^3\bigg),
\end{gather*}
we find that
\begin{gather*}
u = -\frac{x^2}{2}+\frac{y^2}{2}-\frac{x^3}{6}+\frac{x^2y}{2}-\frac{xy^2}{2}+\frac{y^3}{6}.
\end{gather*}
Under the change of variables $(x,y,u)\to(-x,-y,-u)$ we get
\begin{gather}\label{eq:Cayley}
u = \frac{x^2}{2}-\frac{y^2}{2}+\frac{x^3}{6}-\frac{x^2y}{2}+\frac{xy^2}{2}-\frac{y^3}{6}.
\end{gather}
This surface is equivalent to the Cayley surface $u(\widetilde{x},\widetilde{y})=\widetilde{x}\widetilde{y} - \frac{1}{3}\widetilde{x}^3$~\cite{NS-1994}, under the change of~vari\-ables $\widetilde{x} = \frac{1}{\sqrt[3]{2}}(y-x)$, $\widetilde{y} = -\frac{1}{\sqrt[3]{2^2}}(x+y)$.

Finally, one can also verify that~\eqref{eq:Cayley} is a solution to the system of partial differential equations
\begin{gather*}
C_1 + C_2\sqrt{|h|}=A_1=A_2=A_3=0,
\end{gather*}
which the surface must satisfy to be in branch H.3.2.
\end{Example}

\begin{Remark}We note how the recurrence formula and the equivariant moving frame calculus facilitated the above computations by providing us with the matrix $\bm{\mu}^*$ in~\eqref{bmu} essentially for~free.
\end{Remark}

\section{Result summary}\label{sec:summary}

In this section we summarize the results obtained in this paper by listing the normal forms of~surfaces, given as graphs of functions $u(x,y)$, for the different, suitably generic, branches of the equivalence problem we considered in this paper. We~also provide the possible dimensions of~the~self-symmetry group and recall the branches whose differential invariant algebra is gene\-ra\-ted by a single invariant. Note that we do not identify all possible equivalence classes. For~homo\-geneous surfaces, this would require a thorough inspection of all possible constant values that differential invariant can take. For surfaces admitting non-trivial invariants, this would require a detailed analysis of the signature manifold~\cite{O-1995}. Throughout, $\epsilon=\pm 1$, with $\epsilon=1$ for elliptic points and $\epsilon=-1$ for hyperbolic points.

{\it Case} EH.1:
\begin{gather*}
u(x,y)=\frac{1}{2}x^2+\epsilon\frac{1}{2}y^2+\frac{1}{6}x^3+\epsilon\frac{1}{2}x^2y
+\sum_{i,j\geq 0}c_{i(4+j)}\frac{1}{i!(4+j)!}x^iy^{4+j}
+\sum_{\substack{ i+j\geq 4 \\ j<4}}F_{ij}(\mathbf{c})\frac{1}{i!j!}x^iy^j,
\end{gather*}
where $\mathbf{c}$ is the infinite vector of coefficients $c_{i(4+j)}$, $i,j\geq0$ and $F_{ij}$ are certain universal, determinable, functions thereof. These surfaces have self-symmetry groups of dimension~0,~1 or~2, depending on the particularities of $\mathbf{c}$. Also, the algebra of differential invariants is generated by a single fourth order invariant.

{\it Case} EH.2.1:
\begin{gather*}
u(x,y)=\frac{1}{2}x^2+\epsilon\frac{1}{2}y^2+\frac{3\epsilon}{4!}x^4
+\frac{1}{4}x^2y^2+\frac{3\epsilon}{4!}y^4+ \text{h.o.t.},
\end{gather*}
where h.o.t.\ are higher order terms. These surfaces have self-symmetry group of dimension 3, and there are no differential invariants.

{\it Case} H.2.2:
\begin{gather*}
u(x,y)=\frac{1}{2}x^2+\epsilon\frac{1}{2}y^2.
\end{gather*}

These surfaces have self-symmetry groups of dimension 4, and there are no differential inva\-riants.

{\it Case} H.3.1:
\begin{gather*}
u(x,y)=\frac{1}{2}x^2-\frac{1}{2}y^2+\frac{1}{6}x^3-\frac{1}{2}x^2y -\tilde{\epsilon}\frac{1}{2}xy^2+\tilde{\epsilon}\frac{1}{6}y^3
\\ \hphantom{u(x,y)=}
{}+\sum_{i,j\geq 0}c_{(2+i)(2+j)}\frac{1}{(2+i)!(2+j)!}x^{2+i}y^{2+j}
+\sum_{\substack{ i+j\geq 4 \\ j<2~\text{or } i<2}}F_{ij}(\mathbf{c})\frac{1}{i!j!}x^iy^j,
\end{gather*}
where $\mathbf{c}$ is the infinite vector of $c_{(2+i)(2+j)}$, $i,j\geq0$, $c_{22}\neq0$, and $F_{ij}$ are certain universal, determinable, functions thereof. These surfaces have self-symmetry groups of dimension~0,~1 or~2, depending on the particularities of~$\mathbf{c}$. Furthermore, the algebra of differential invariants is generated by a single fourth order invariant.

{\it Case} H.3.2:
\begin{gather*}
u(x,y)=\frac{1}{2}x^2-\frac{1}{2}y^2+\frac{1}{6}x^3-\frac{1}{2}x^2y+\frac{1}{2}xy^2-\frac{1}{6}y^3.
\end{gather*}

The self-symmetry group has dimension 3, and there are no differential invariants.

{\it Case} P1.1.1:
\begin{gather*}
u(x,y)=\frac{1}{2}x^2+\frac{1}{2}x^2y+\frac{1}{6}x^3y+\frac{1}{2}x^2y^2
+\sum_{i,j\geq 0}c_{(5+i)j}\frac{1}{(5+i)!j!}x^{5+i}y^{j}
+\!\!\sum_{\substack{ i+j\geq 5 \\ i<5}}\!F_{ij}(\mathbf{c})\frac{1}{i!j!}x^iy^j,
\end{gather*}
where $\mathbf{c}$ is the infinite vector of $c_{(5+i)(j)}$, $i,j\geq0$ and $F_{ij}$ are certain universal, determinable, functions thereof. These surfaces have self-symmetry groups of dimension 0, 1 or~2, depending on the particularities of $\mathbf{c}$. In~this case, the algebra of differential invariants is generated by a~fifth order invariant.

{\it Case} P.1.1.2.1:
\begin{gather*}
u(x,y) = \frac{1}{2}x^2+\frac{1}{2}x^2y+\frac{1}{2}x^2y^2+\frac{1}{5!}x^5+\frac{1}{2}x^2y^3+\frac{1}{2}x^2y^4 + \frac{1}{30} x^5y + \text{h.o.t.},
\end{gather*}
where h.o.t.\ are higher order terms. The self-symmetry group has dimension~2, and the diffe\-ren\-tial invariant algebra is generated by a seventh order invariant.

{\it Case} P.1.1.2.2:
\begin{gather*}
u(x,y)=\frac{1}{2}x^2+\frac{1}{2}x^2y+\frac{1}{2}x^2y^2+\frac{1}{5!}x^5+\frac{1}{2}x^2y^3+\frac{1}{2}x^2y^4 +\text{h.o.t.},
\end{gather*}
where h.o.t.\ are higher order terms. The self-symmetry group has dimension 4, and there are no differential invariants.

{\it Case} P.1.2.1:
\begin{gather*}
u(x,y)=\frac{1}{2}x^2+\frac{1}{4!}x^4+\sum_{i\geq 0}c_{(5+i)0}\frac{1}{(5+i)!}x^{5+i}.
\end{gather*}
The self-symmetry group has dimension 3, 4 or 5 depending on the series of $c_{(5+i)0}$, and the invariant differential algebra is generated by a fifth invariant.

{\it Case} P.1.2.2:
\begin{gather*}
u(x,y)=\frac{1}{2}x^2.
\end{gather*}
The symmetry group has dimension 6, and there are no differential invariants.

{\it Case} P.2:
\begin{gather*}
u(x,y)=0
\end{gather*}
has a 9-dimensional self-symmetry group, and there are no differential invariants.

\subsection*{Acknowledgement}

We would like to thank the referees for their valuable comments, which helped improve the exposition of the paper.

\pdfbookmark[1]{References}{ref}
\LastPageEnding


\begin{thebibliography}{99}
\footnotesize\itemsep=-.5pt

\bibitem{ADV-1997}
Abdalla B.E., Dillen F., Vrancken L., Affine homogeneous surfaces in {${\mathbb
 R}^3$} with vanishing {P}ick invariant, \href{https://doi.org/10.1007/BF02940821}{\textit{Abh. Math. Sem. Univ.
 Hamburg}} \textbf{67} (1997), 105--115.

\bibitem{B-1923}
Blaschke W., Vorlesungen \"uber {G}eometrie und {G}eometrische {G}rundlagen von
 {E}insteins {R}elativit\"atstheorie {II}: {A}ffine {D}ifferentialgeometrie,
 \href{https://doi.org/10.1007/978-3-642-47392-0}{Verlag Von Julius Springer}, Berlin, 1923.

\bibitem{CM-2020}
Chen Z., Merker J., On differential invariants of parabolic surfaces,
 \href{https://arxiv.org/abs/1908.07867}{arXiv:1908.07867}.

\bibitem{DMMSV-1991}
Dillen F., Mart\'{\i}nez A., Mil\'an F., Garcia~Santos F., Vrancken L., On the
 {P}ick invariant, the affine mean curvature and the {G}auss curvature of
 affine surfaces, \href{https://doi.org/10.1007/BF03323199}{\textit{Results Math.}} \textbf{20} (1991), 622--642.

\bibitem{DKR-1996}
Doubrov B., Komrakov B., Rabinovich M., Homogeneous surfaces in the
 three-dimensional affine geometry, in Geometry and Topology of Submanifolds,
 {VIII} ({B}russels, 1995/{N}ordfjordeid, 1995), \href{https://doi.org/10.1142/9789814530873}{World Sci. Publ.}, River Edge,
 NJ, 1996, 168--178.

\bibitem{DMT-2020}
Doubrov B., Merker J., The D., Classification of simply-transitive Levi
 non-degenerate hypersurfaces in $\mathbb{C}^3$, \href{https://arxiv.org/abs/2010.06334}{arXiv:2010.06334}.

\bibitem{EE-1999}
Eastwood M., Ezhov V., On affine normal forms and a classification of
 homogeneous surfaces in affine three-space, \href{https://doi.org/10.1023/A:1005083518793}{\textit{Geom. Dedicata}}
 \textbf{77} (1999), 11--69.

\bibitem{FO-1999}
Fels M., Olver P.J., Moving coframes. {II}. {R}egularization and theoretical
 foundations, \href{https://doi.org/10.1023/A:1006195823000}{\textit{Acta Appl. Math.}} \textbf{55} (1999), 127--208.

\bibitem{G-74}
Griffiths P., On {C}artan's method of {L}ie groups and moving frames as applied
 to uniqueness and existence questions in differential geometry, \href{https://doi.org/10.1215/S0012-7094-74-04180-5}{\textit{Duke
 Math.~J.}} \textbf{41} (1974), 775--814.

\bibitem{G-1963}
Guggenheimer H.W., Differential geometry, McGraw-Hill Book Co., Inc., New
 York~-- San Francisco~-- Toronto~-- London, 1963.

\bibitem{HO-2007}
Hubert E., Olver P.J., Differential invariants of conformal and projective
 surfaces, \href{https://doi.org/10.3842/SIGMA.2007.097}{\textit{SIGMA}} \textbf{3} (2007), 097, 15~pages,
 \href{https://arxiv.org/abs/0710.0519}{arXiv:0710.0519}.

\bibitem{J-1977}
Jensen G.R., Higher order contact of submanifolds of homogeneous spaces,
 \textit{Lecture Notes in Math.}, Vol.~610, \href{https://doi.org/10.1007/BFb0068415}{Springer-Verlag}, Berlin~-- New
 York, 1977.

\bibitem{KO-2003}
Kogan I.A., Olver P.J., Invariant {E}uler--{L}agrange equations and the
 invariant variational bicomplex, \href{https://doi.org/10.1023/A:1022993616247}{\textit{Acta Appl. Math.}} \textbf{76}
 (2003), 137--193.

\bibitem{KL-2006}
Kruglikov B., Lychagin V., Invariants of pseudogroup actions: homological
 methods and finiteness theorem, \href{https://doi.org/10.1142/S0219887806001478}{\textit{Int.~J. Geom. Methods Mod. Phys.}}
 \textbf{3} (2006), 1131--1165, \href{https://arxiv.org/abs/math.DG/0511711}{arXiv:math.DG/0511711}.

\bibitem{KL-2016}
Kruglikov B., Lychagin V., Global {L}ie--{T}resse theorem, \href{https://doi.org/10.1007/s00029-015-0220-z}{\textit{Selecta
 Math. (N.S.)}} \textbf{22} (2016), 1357--1411, \href{https://arxiv.org/abs/1111.5480}{arXiv:1111.5480}.

\bibitem{L-1893}
Lie S., Scheffers G., Vorlesungen \"uber continuierliche {G}ruppen mit
 {G}eometrischen und anderen {A}nwendungen, B.G.~Teubner, Leipzig, 1893.

\bibitem{M-2010}
Mansfield E.L., A practical guide to the invariant calculus, \textit{Cambridge
 Monographs on Applied and Computational Mathematics}, Vol.~26, \href{https://doi.org/10.1017/CBO9780511844621}{Cambridge
 University Press}, Cambridge, 2010.

\bibitem{M-2019}
Merker J., Affine rigidity without integration, \href{https://arxiv.org/abs/1903.00889}{arXiv:1903.00889}.

\bibitem{NS-1994}
Nomizu K., Sasaki T., Affine differential geometry, \textit{Cambridge Tracts in
 Mathematics}, Vol.~111, Cambridge University Press, Cambridge, 1994.

\bibitem{O-1995}
Olver P.J., Equivalence, invariants, and symmetry, \href{https://doi.org/10.1017/CBO9780511609565}{Cambridge University Press},
 Cambridge, 1995.

\bibitem{O-2001}
Olver P.J., Moving frames: a brief survey, in Symmetry and Perturbation Theory,
 Editors D.~Bambusi, G.~Gaeta, M.~Cadoni, \href{https://doi.org/10.1142/9789812794543_0020}{World Sci.}, Singapore, 2001,
 143--150.

\bibitem{O-2007}
Olver P.J., Generating differential invariants, \href{https://doi.org/10.1016/j.jmaa.2006.12.029}{\textit{J.~Math. Anal. Appl.}}
 \textbf{333} (2007), 450--471.

\bibitem{O-2009diff}
Olver P.J., Differential invariants of surfaces, \href{https://doi.org/10.1016/j.difgeo.2008.06.020}{\textit{Differential Geom.
 Appl.}} \textbf{27} (2009), 230--239.

\bibitem{O-2011}
Olver P.J., Recursive moving frames, \href{https://doi.org/10.1007/s00025-011-0153-6}{\textit{Results Math.}} \textbf{60} (2011),
 423--452.

\bibitem{O-2016}
Olver P.J., Equivariant moving frames for Euclidean surfaces, {P}reprint,
 University of Minnesota, 2016, available at
 \url{https://www-users.math.umn.edu/~olver/mf_/eus.pdf}.

\bibitem{OP-2007}
Olver P.J., Pohjanpelto J., Differential invariants for {L}ie pseudo-groups, in
 Gr\"obner bases in symbolic analysis, \textit{Radon Ser. Comput. Appl.
 Math.}, Vol.~2, \href{https://doi.org/10.1515/9783110922752}{Walter de Gruyter}, Berlin, 2007, 217--243.

\bibitem{SS-1962}
Schirokow P.A., Schirokow A.P., Affine {D}ifferentialgeometrie, B.G.~Teubner
 Verlagsgesellschaft, Leipzig, 1962.

\bibitem{S-2010}
Seiler W.M., Involution: the formal theory of differential equations and its
 applications in computer algebra, \textit{Algorithms and Computation in
 Mathematics}, Vol.~24, \href{https://doi.org/10.1007/978-3-642-01287-7}{Springer-Verlag}, Berlin, 2010.

\end{thebibliography}
\end{document}